\documentclass[sigconf,anonymous=false]{acmart}
\AtBeginDocument{%
  \providecommand\BibTeX{{%
    \normalfont B\kern-0.5em{\scshape i\kern-0.25em b}\kern-0.8em\TeX}}}

\setcopyright{acmcopyright}
\copyrightyear{2022}
\acmYear{2022}
\copyrightyear{2022}
\acmYear{2022}
\setcopyright{acmlicensed}\acmConference[WSDM '22]{Proceedings of the
Fifteenth ACM International Conference on Web Search and Data
Mining}{February 21--25, 2022}{Tempe, AZ, USA}
\acmBooktitle{Proceedings of the Fifteenth ACM International Conference on
Web Search and Data Mining (WSDM '22), February 21--25, 2022, Tempe, AZ,
USA}
\acmPrice{15.00}
\acmDOI{10.1145/3488560.3498490}
\acmISBN{978-1-4503-9132-0/22/02}
\acmSubmissionID{637}

\usepackage{layouts}

\usepackage[nolabel]{showlabels} % [nolabel]
\usepackage[titles]{tocloft}
\usepackage{amsmath}
\usepackage{savesym}
\savesymbol{iint}
\usepackage{amsthm}

\usepackage{empheq}
\usepackage{physics} %\bra \ket \braket{...}
\usepackage[inter-unit-product =\cdot]{siunitx}
\restoresymbol{TXF}{iint}
\usepackage[super]{nth}

\usepackage{mathtools}

\usepackage{csquotes}

\usepackage{bold-extra} % allows combining boldface and smallcaps
\usepackage{enumerate}
\usepackage{enumitem} % pour utiliser description
\setlist[1]{leftmargin=*} % < Usually a good idea
\usepackage{color}
\definecolor{mygreen}{rgb}{0,0.6,0}
\definecolor{mygray}{rgb}{0.5,0.5,0.5}
\definecolor{mymauve}{rgb}{0.58,0,0.82}
\usepackage{xspace}
\usepackage{datetime}

\usepackage{dsfont}

\usepackage{bm}

\usepackage{eurosym}

\usepackage{numprint}

\usepackage{lettrine}

\usepackage{lipsum}

\usepackage{accents}

\usepackage{marginnote}
\usepackage{comment}

\usepackage{dirtytalk}

\usepackage[font=footnotesize,labelfont=bf]{caption}
\usepackage{subcaption}

\usepackage[numbered]{bookmark}

\usepackage{mathtools}
\mathtoolsset{showonlyrefs=false} % Fisher's rule of equation numbering

\usepackage{pgfplots}
\pgfplotsset{compat=newest}
\usepgfplotslibrary{groupplots}
\usepgfplotslibrary{dateplot}

\usepackage{tikz}
\usetikzlibrary{patterns}
\usetikzlibrary{shapes}
\usetikzlibrary{calc}
\usetikzlibrary{decorations.pathreplacing,calligraphy}

\usepackage[bottom]{footmisc}

\usepackage{fancyhdr}
\fancypagestyle{plain}{%
	\fancyhf{}% Clear header and footer
	\fancyhf[FLO,FRE]{\thepage}
	\fancyhf[FLE,FRO]{\hyperlink{toc}{\footnotesize$\uparrow$\texttt{TOC}}}
}

\usepackage[Lenny]{fncychap}
\usepackage{epigraph}

\usepackage[symbols,acronym,section]{glossaries} % toc
\usepackage{multicol}

\usepackage{pgfornament}

\usepackage{import}

\usepackage[title,toc,header]{appendix} % [title,toc,page,header,titletoc]

\usepackage{algpseudocode}
\usepackage{algorithm}
\algdef{SE}[DOWHILE]{Do}{doWhile}{\algorithmicdo}[1]{\algorithmicwhile\ #1}% custom "do ... while ... command"
\usepackage{listings}
\usepackage{listingsutf8}

\newcommand{\ixitem}{i}  % index of items
\newcommand{\ixrank}{k}  % index of rank
\newcommand{\ndoc}{n}

\newcommand{\query}{q}
\newcommand{\expo}{\Pi}
\newcommand{\zone}{Z}
\newcommand{\argsort}{\texttt{argsort}}

\newcommand{\temperature}{\tau}

\newcommand{\e}[1]{\cdot\!10^{#1}}

\newcommand{\N}{\mathbb{N}}

\newcommand{\R}{\mathbb{R}}

\newcommand{\cA}{\mathcal{A}}

\newcommand{\cD}{\mathcal{D}}

\newcommand{\cI}{\mathcal{I}}

\newcommand{\cS}{\mathcal{S}}

\DeclareMathAlphabet\mathbfcal{OMS}{cmsy}{b}{n}
\newcommand{\latinvector}[1]{\mathbf{#1}}
\renewcommand{\va}{\latinvector{a}}
\renewcommand{\vb}{\latinvector{b}}

\newcommand{\vp}{\latinvector{p}}

\renewcommand{\vv}{\latinvector{v}}

\newcommand{\vx}{\latinvector{x}}

\newcommand{\vone}{\mathbf{1}}
\newcommand{\vzero}{\mathbf{0}}
\newcommand{\vgamma}{\bm{\gamma}}
\newcommand{\vrho}{\bm{\rho}}

\newcommand{\vnu}{\bm{\nu}}
\newcommand{\vphi}{\bm{\phi}}
\newcommand{\vcE}{\mathbfcal{E}}
\newcommand{\vcEt}{\mathbfcal{E}^*}

\newcommand{\defeq}{\vcentcolon=}

\DeclareMathOperator{\Conv}{Conv}
\DeclareMathOperator{\logtwo}{log_2}
\theoremstyle{plain}

\theoremstyle{remark}

\theoremstyle{plain}\newtheorem{theo}{Theorem}

\theoremstyle{remark}

\theoremstyle{definition}\newtheorem{defi}{Definition}

\theoremstyle{plain}

\theoremstyle{plain}

\newcommand{\fdem}{\hspace*{\fill}~$\Box$\par\endtrivlist\unskip}

{
   \end{minipage}
   \vspace*{\stretch{3}}
}

\newcommand{\seq}[2]{\{#1,\hdots,#2\}}

\providecommand{\keywords}[1]
{
	\small
	\textbf{\textit{Keywords---}} #1
}

\newacronym{acm}{ACM}{Association for Computing Machinery}
\newacronym{adaboost}{AdaBoost}{Adaptative Boosting}
\newacronym{ads}{ADS}{Automated Decision System}
\newacronym{ai}{AI}{Artificial Intelligence}
\newacronym{aif}{AIF}{Average Individual Fairness}
\newacronym{aka}{a.k.a.}{also known as}
\newacronym{ama}{AMA}{Artificial Moral Agent}
\newacronym{ap}{AP}{Average Precision}
\newacronym{bw}{BW}{Balanced Words}
\newacronym{bert}{BERT}{Bidirectional Encoder Representations from Transformers}
\newacronym{bisg}{BISG}{Bayesian Improved Surname Geocoding}
\newacronym{bm}{BM}{Best Matching}
\newacronym{bo}{BO}{Bayesian Optimisation}
\newacronym{bpr}{BPR}{Bayesian Personalized Ranking}
\newacronym{bp}{BP}{Belief Propagation}
\newacronym{bvn}{BvN}{Birkhoff-von Neumann}
\newacronym{cb}{CB}{Contextual Bandits}
\newacronym{cca}{CCA}{Canonical Correlation Analysis}
\newacronym{cdssm}{C-DSSM}{Convolutional Deep Structured Semantic Model}
\newacronym{cf}{CF}{Collaborative Filtering}
\newacronym{cf.}{cf.}{confer}
\newacronym{clara}{CLARA}{CLick And RAnking}
\newacronym{clef}{CLEF}{Conference and Labs of the Evaluation Forum}
\newacronym{clm}{CLM}{Cumulative Link Model}
\newacronym{clsm}{CLSM}{Convolutional Latent Semantic Model}
\newacronym{cm}{CM}{Cascade Model}
\newacronym{cnn}{CNN}{Convolutional Neural Network}
\newacronym{compas}{COMPAS}{Correctional Offender Management Profiling for Alternative Sanctions}
\newacronym{cpu}{CPU}{Central Processing Unit}
\newacronym{crs}{CRS}{Conversational Recommender System}
\newacronym{csc}{CSC}{Cost Sensitive Classification}
\newacronym{csi}{CSI}{Comité de suivi de thèse}
\newacronym{css}{CSS}{Complementarity Sum Sampling}
\newacronym{ctr}{CTR}{Click Through Rate}
\newacronym{ctrl}{Ctrl}{Controller}
\newacronym{da}{DA}{Data Augmentation}
\newacronym{dag}{DAG}{Directed Acyclic Graph}
\newacronym{dbn}{DBN}{Dynamic Bayesian Network}
\newacronym{dcg}{DCG}{Discounted Cumulative Gain}
\newacronym{dcm}{DCM}{Discrete Choice Model}
\newacronym{dct}{DCT}{Discrete Cosine Transform}
\newacronym{dlrm}{DLRM}{Deep Learning Recommendation Model}
\newacronym{dm}{DM}{Decision Maker}
\newacronym{dnn}{DNN}{Deep Neural Network}
\newacronym{dp}{DP}{Differential Privacy}
\newacronym{dssm}{DSSM}{Deep Structured Semantic Model}
\newacronym{ebm}{EBM}{Energy-Based Model}
\newacronym{ebms}{EBMs}{Energy-Based Models}
\newacronym{ece}{ECE}{Empirical Calibration Error}
\newacronym{ef}{EF}{Envy-Free}
\newacronym{ef1}{EF1}{Envy-Free up to one item}
\newacronym{efr}{EFR}{Expected First Relevant}
\newacronym{eg}{e.g.}{exempli gratia}
\newacronym{elmo}{ELMo}{Embeddings from Language Models}
\newacronym{erm}{ERM}{Empirical Risk Minimization}
\newacronym{err}{ERR}{Expected Reciprocal Rank}
\newacronym{ers}{ERS}{Empirical Relational System}
\newacronym{etc}{etc.}{et cetera}
\newacronym{expo}{Expo}{Expohedron}
\newacronym{fair}{FAIR}{Fairness-Aware Information Retrieval}
\newacronym{fire}{FIRE}{Forum for Information Retrieval Evaluation}
\newacronym{fm}{FM}{Factorization Machine}
\newacronym{fultr}{FULTR}{Fair Un-biased Learning-to-Rank}
\newacronym{gan}{GAN}{Generative Adversarial Network}
\newacronym{gbdt}{GBDT}{Gradient Boosted Decision Tree}
\newacronym{gcn}{GCN}{Graph Convolution Network}
\newacronym{gdpr}{GDPR}{General Data Protection Regulation}
\newacronym{gls}{GLS}{Grötschel, Lovász and Schrijver}
\newacronym{gnn}{GNN}{Graph Neural Network}
\newacronym{gpt}{GPT}{Generative Pre-trained Transformer}
\newacronym{gpu}{GPU}{Graphics Processing Unit}
\newacronym{hpc}{HPC}{High Performance Computing}
\newacronym{hr}{HR}{Hit Rate}
\newacronym{idcg}{IDCG}{Ideal Discounted Cumulative Gain}
\newacronym{idf}{IDF}{Inverse Document Frequency}
\newacronym{ie}{i.e.,}{id est}
\newacronym{iid}{i.i.d.}{independent and identically distributed}
\newacronym{ilp}{ILP}{Integer Linear Program}
\newacronym{ipm}{IPM}{Integral Probability Metric}
\newacronym{ips}{IPS}{Inverse Propensity Score}
\newacronym{ipu}{IPU}{Intelligence Processing Unit}
\newacronym{ir}{IR}{Information Retrieval}
\newacronym{irgan}{IRGAN}{Information Retrieval Generative Adversarial Network}
\newacronym{irm}{IRM}{Invariant Risk Minimization}
\newacronym{irsvm}{IR-SVM}{Information Retrieval Support Vector Machine}
\newacronym{kl}{KL}{Kullback-Leibler}
\newacronym{knn}{KNN}{$K$-nearest neighbors}
\newacronym{lds}{LDS}{Low Discrepancy Sequence}
\newacronym{lc}{LC}{Latent Cross}
\newacronym{letor}{LETOR}{LEarning TO Rank for Information Retrieval}
\newacronym{lm}{LM}{Language Model}
\newacronym{lp}{LP}{Linear Program}
\newacronym{lsi}{LSI}{Latent Semantic Indexing}
\newacronym{lspi}{LSPI}{Least Squares Policy Iteration}
\newacronym{lstdq}{LSTD-Q}{Least Squares Temporal Difference Q-learning}
\newacronym{lte}{LTE}{Learning To Expose}
\newacronym{ltr}{LTR}{Learning to Rank}
\newacronym{mab}{MAB}{Multi-Armed Bandit}
\newacronym{map}{MAP}{Mean Average Precision}
\newacronym{mcrank}{McRank}{Multi-class Classification for Ranking}
\newacronym{mdp}{MDP}{Markov Decision Process}
\newacronym{me}{ME}{Maximum Entropy}
\newacronym{mf}{MF}{Matrix Factorization}
\newacronym{mfr}{MFR}{Mean First Relevant}
\newacronym{mhr}{MHR}{Multiple Hyperplane Ranker}
\newacronym{mi}{MI}{Mutual Information}
\newacronym{miai}{MIAI}{Multidisciplinary Institute for AI}
\newacronym{milp}{MILP}{Mixed Integer Linear Program}
\newacronym{mimd}{MIMD}{Multiple Instructions Multiple Data}
\newacronym{ml}{ML}{Machine Learning}
\newacronym{mlo}{MLO}{Machine Learning and Optimization}
\newacronym{mlp}{MLP}{MultiLayer Perceptron}
\newacronym{mms}{MMS}{MaxiMin Share}
\newacronym{mnist}{MNIST}{Modified National Institute of Standards and Technology}
\newacronym{molp}{MOLP}{Multi-Objective Linear Program}
\newacronym{moo}{MOO}{Multi-Objective Optimisation}
\newacronym{mpt}{MPT}{Modern Portfolio Theory}
\newacronym{mrr}{MRR}{Mean Reciprocal Rank}
\newacronym{mrs}{MRS}{Mulstistakeholder Recommender System}
\newacronym{mslr}{MSLR}{Microsoft Learning to Rank}
\newacronym{msp}{MSP}{Multi-Sided Platform}
\newacronym{mvd}{MVD}{MultiValued Dependency}
\newacronym{ndcg}{nDCG}{Normalized Discounted Cumulative Gain}
\newacronym{ndf}{NDF}{Nearest Driver First}
\newacronym{ngcf}{NGCF}{Neural Graph Collaborative Filtering}
\newacronym{nle}{NLE}{Naver Labs Europe}
\newacronym{nlp}{NLP}{Natural Language Processing}
\newacronym{nmf}{NMF}{Non-negative matrix factorization}
\newacronym{nn}{NN}{Neural Network}
\newacronym{nrs}{NRS}{Numerical Relational System}
\newacronym{ntcir}{NTCIR}{NII Testbeds and Community for Information Access Research}
\newacronym{ood}{OOD}{Out-Of-Distribution}
\newacronym{pb}{PB}{Poisson Binomial distribution}
\newacronym{pm}{PM}{Poisson Multinomial distribution}
\newacronym{pbm}{PBM}{Position Based Model}
\newacronym{pdp}{PDP}{Pigou-Dalton Principle}
\newacronym{pf}{PF}{Poisson Factorization}
\newacronym{pl}{PL}{Plackett-Luce}
\newacronym{pmf}{pmf}{probability mass function}
\newacronym{pomdp}{POMDP}{Partially Observed Markov Decision Process}
\newacronym{prairie}{PRAIRIE}{PaRis Artificial Intelligence Research InstitutE}
\newacronym{pranking}{PRanking}{Perceptron based Ranking}
\newacronym{pfs}{PFS}{Proportional Fair Share}
\newacronym{prp}{PRP}{Probability Ranking Principle}
\newacronym{qp}{QP}{Quadratic Program}
\newacronym{rbm}{RBM}{Restricted Boltzmann Machine}
\newacronym{rbp}{RBP}{Rank Biased Precision}
\newacronym{rcm}{RCM}{Random Click Model}
\newacronym{rct}{RCT}{Randomized Controlled Trial}
\newacronym{rc}{RC}{Rank Correlation}
\newacronym{relu}{ReLU}{Rectified Linear Unit}
\newacronym{resp}{resp.}{respectively}
\newacronym{rl}{RL}{Reinforcment Learning}
\newacronym{rnn}{RNN}{Recurrent Neural Network}
\newacronym{roc}{ROC}{Receiver Operating Characteristic}
\newacronym{rr}{RR}{Round-Robin}
\newacronym{rs}{RS}{Recommender System}
\newacronym{rv}{r.v.}{random variable}
\newacronym{sa}{SA}{Self Attention}
\newacronym{sar}{SaR}{Search and Recommendation}
\newacronym{scm}{SCM}{Structural Causal Model}
\newacronym{sea}{SEA}{Safe Exploration Algorithm}
\newacronym{serp}{SERP}{Search Engine Result Page}
\newacronym{sgd}{SGD}{Stochastic Gradient Descent}
\newacronym{sigmod}{SIGMOD}{Special Interest Group on Management of Data}
\newacronym{slim}{SLiM}{Sparse Linear Model}
\newacronym{st}{s.t.}{such that}
\newacronym{svd}{SVD}{Singular Value Decomposition}
\newacronym{svm}{SVM}{Support Vector Machine}
\newacronym{svmmap}{SVM$^{map}$}{Support Vector Machine maximum average precision}
\newacronym{tf}{TF}{Term Frequency}
\newacronym{tgn}{TGN}{Temporal Graph Network}
\newacronym{trec}{TREC}{Text REtrieval Conference}
\newacronym{tsp}{TSP}{Traveling Salesman Problem}
\newacronym{ubm}{UBM}{User Browsing Model}
\newacronym{ui}{UI}{User Interface}
\newacronym{url}{URL}{Uniform Resource Locator}
\newacronym{vae}{VAE}{Variational AutoEncoder}
\newacronym{vi}{VI}{Variational Inequality}
\newacronym{vsm}{VSM}{Vector Space Model}
\newacronym{wdf}{WDF}{Worst-off Driver First}
\newacronym{wlog}{w.l.o.g.}{without loss of generality}
\newacronym{wrt}{w.r.t.}{with respect to}
\newacronym{www}{www}{world wide web}

\newglossaryentry{adarank}{name=AdaRank, description={}}
\newglossaryentry{apprank}{name=AppRank, description={}}
\newglossaryentry{bn-decomp}{name=Birkhoff-von Neumann decomposition, description={A (generally non-unique) decomposition of a doubly stochastic matrix into a convex sum of permutation matrices}}
\newglossaryentry{boosting}{name=Boosting, description={Using weighed prediction of a ton of weak learners to build One strong learner, and rule them all.}}
\newglossaryentry{borda-count}{name=Borda Count, description={A vote aggregation method formalized by the chevalier de Borda}}
\newglossaryentry{jm-smooth}{name=Jelinek-Mercer Smoothing, description={Interpolating with the prior (convex combination)}}
\newglossaryentry{katz}{name={Katz measure}, description={A measure to quantify the \emph{centrality} of a node in a graph}}
\newglossaryentry{listmle}{name=ListMLE, description={}}
\newglossaryentry{listnet}{name=ListNet, description={}}
\newglossaryentry{lstm}{name=Long Short-Term Memory (LSTM), description={Memory cell structure in a \acrlong{nn} \cite{hochreiter_long_1997}. See also \url{https://en.wikipedia.org/wiki/Long_short-term_memory}}}
\newglossaryentry{luce}{name=Luce, description={A probability distribution on permutations}}
\newglossaryentry{oracle-efficient}{name=Oracle-efficient algorithm, description={An algorithm that is given access to any standard, fairness-free learning heuristic, from \cite{kearns_average_2019}}}
\newglossaryentry{ord-reg}{name=Ordinal Regression, description={Regression where the target space is characterized only by the relative order of its elements}}
\newglossaryentry{permurank}{name=PermuRank, description={}}
\newglossaryentry{rankgp}{name=RankGP, description={A genetic algorithm for learning rankings}}
\newglossaryentry{rank-agg}{name=Rank Aggregation, description={The problem of finding the ranking with maximum agreement to (possibly incompatible) pairwise orderings. It has been proven to be NP-hard.}}
\newglossaryentry{rank-boost}{name=RankBoost, description={A Boosting method using AdaBoost \cite{freund_efficient_2003}}}
\newglossaryentry{ranknet}{name=RankNet, description={A learning-to-rank algorithm \cite{burges_learning_2005}}}
\newglossaryentry{reciprocal-recommender}{name={Reciprocal Recommender}, description={A recommender platform where two types of users are recommended to each other heterosexually, \acrshort{eg} workerss and employers, men and women, french russian-learners and russian french-learners}}
\newglossaryentry{shilling}{name={Shilling Attack}, description={When an item producer gives himself lots of high ratings}}
\newglossaryentry{softrank}{name=SoftRank, description={}}
\newglossaryentry{vc-dim}{name=Vapnik-Chervonenkis-dimension, description={cardinality of the largest set of points that the algorithm can shatter, from wiki}}
\newglossaryentry{vertical}{name=vertical, description={Within a websearch, the results are presented in blocks of e.g. videos, news articles, websites, pictures. Those blocks are called verticals}}

\usepackage{balance}

\begin{document}
% \fancyhead{}

\title{Introducing the Expohedron for Efficient Pareto-optimal Fairness-Utility Amortizations in Repeated Rankings}

\author{Till Kletti}
\email{till.kletti@naverlabs.com}
\orcid{0000-0002-8853-4618}
\affiliation{%
  \institution{Naver Labs Europe}
  \streetaddress{6 chemin de Maupertuis}
%   \city{Meylan}
  \country{France}
  \postcode{38700}
}

\author{Jean-Michel Renders}
\email{jean-michel.renders@naverlabs.com}
\orcid{0000-0002-7516-3707}
\affiliation{%
  \institution{Naver Labs Europe}
  \streetaddress{6 chemin de Maupertuis}
%   \city{Meylan}
  \country{France}
  \postcode{38700}
}

% \author{Sihem Amer-Yahia}
% \email{sihem.amer-yahia@univ-grenoble-alpes.fr}
% \affiliation{%
%   \institution{UGA}
%   \city{Saint-Martin-d'Hères}
%   \country{France}}

\author{Patrick Loiseau}
\email{patrick.loiseau@inria.fr}
\affiliation{%
  \institution{Univ. Grenoble Alpes, Inria, CNRS, Grenoble INP, LIG}
%   \city{Grenoble}
  \country{France}
}

\renewcommand{\shortauthors}{Kletti et al.}

\begin{abstract}
  % Patrickly influenced

We consider the problem of computing a sequence of rankings that maximizes consumer-side utility while minimizing producer-side individual unfairness of exposure.
While prior work has addressed this problem using linear or quadratic programs on bistochastic matrices, such approaches, relying on \acrfull{bvn} decompositions, are too slow to be implemented at large scale.

In this paper we introduce a geometrical object, a polytope that we call \emph{expohedron}, whose points represent all achievable exposures of items for a \acrfull{pbm}.
We exhibit some of its properties and lay out a Carathéodory decomposition algorithm with complexity $O(\ndoc^2\log(\ndoc))$ able to express any point inside the expohedron as a convex sum of at most $\ndoc$ vertices, where $\ndoc$ is the number of items to rank.
Such a decomposition makes it possible to express any feasible target exposure as a distribution over at most $\ndoc$ rankings.
Furthermore we show that we can use this polytope to recover the whole Pareto frontier of the multi-objective fairness-utility optimization problem, using a simple geometrical procedure with complexity $O(\ndoc^2\log(\ndoc))$.
Our approach compares favorably to linear or quadratic programming baselines in terms of algorithmic complexity and empirical runtime and is applicable to any merit that is a non-decreasing function of item relevance.
% to any reasonable notion of merit.
Furthermore our solution can be expressed as a distribution over only $\ndoc$ permutations, instead of the $(\ndoc-1)^2 + 1$ achieved with \acrshort{bvn} decompositions.
We perform experiments on synthetic and real-world datasets, confirming our theoretical results.

\end{abstract}




\begin{teaserfigure}
    \vspace{-\baselineskip}
    \begin{subfigure}[b]{0.3\textwidth}
        \centering
        \begin{tikzpicture}
	[scale=5.000000,
	back/.style={loosely dotted, thin},
	edge/.style={color=black, thick},
	facet/.style={fill=white,fill opacity=0.300000},
	vertex/.style={inner sep=1pt,circle,draw=red!25!black,fill=red!75!black,thick}]
  % Local definitions
  \def\costhirty{0.8660256}

  % Colors
  \colorlet{anglecolor}{green!50!black}
  \colorlet{sincolor}{red}
  \colorlet{tancolor}{orange!80!black}
  \colorlet{coscolor}{blue}

  % Styles
  \tikzstyle{axes}=[]
  \tikzstyle{important line}=[very thick]
  \tikzstyle{information text}=[rounded corners,fill=red!10,inner sep=1ex]

  % The graphic
  % \draw[style=help lines,step=0.1cm] (0,0) grid (1,1);

  % \begin{scope}[style=axes]
  %   \draw[->] (-0.1,0) -- (1.2,0) node[right] {$\mathcal{E}_1$};
  %   \draw[->] (0,-0.1) -- (0,1.2) node[above] {$\mathcal{E}_2$};
	%
  %   \foreach \x/\xtext in {0, 0.25/\frac{1}{4}, 0.5/\frac{1}{2}, 0.75/\frac{3}{4}, 1}
  %     \draw[xshift=\x cm] (0pt,1pt) -- (0pt,-1pt) node[below,fill=white]
  %           {$\xtext$};
	%
  %   \foreach \y/\ytext in {0, 0.25/\frac{1}{4}, 0.5/\frac{1}{2}, 0.75/\frac{3}{4}, 1}
  %     \draw[yshift=\y cm] (1pt,0pt) -- (-1pt,0pt) node[left,fill=white]
  %           {$\ytext$};
  % \end{scope}

	% \draw[color=red] (1, 0.25) -- (0.5, 1);
	% \draw[dashed] (0, 0) -- (1, 1);
	% \draw[dotted] (0, 0) coordinate (O) -- (8/10, 11/20) coordinate (X);
	% \draw[<-] (8/10, 11/20) .. controls (1,0.6) .. (1.1,0.6) node[align=left, right] {$\arg\min||\mathcal{E}||_2$};
	% \draw[<-] (7/10, 7/10) .. controls (0.9,0.7) .. (1.1,3/4) node[align=left, right] {Equity};

	\draw[edge] (1, 0.63093) -- (0.63093, 1);
	\node[vertex] at (1, 0.63093)     {};
	\node[vertex] at (0.63093, 1)     {};

	% \filldraw (1,0.63) coordinate (A) circle (0.5pt) node[align=left, right] {$(\gamma_1,\gamma_2)$} --
	% (0.63,1) circle (0.5pt) node[align=left, above] {$\gamma_2,\gamma_1$};

	% Right angle
	% \draw pic [draw=black,angle eccentricity=1,pic text=$\cdot$,scale=0.3] {right angle = O--X--A};

\end{tikzpicture}
        \caption{$\ndoc=2$. A line segment in $\R^2$}
    \end{subfigure}
    \hfill
    \begin{subfigure}[b]{0.3\textwidth}
        \centering
        \begin{tikzpicture}%
	[x={(-0.707031cm, -0.408259cm)},
	y={(0.707183cm, -0.408200cm)},
	z={(0.000025cm, 0.816516cm)},
	scale=5.000000,
	back/.style={loosely dotted, thin},
	edge/.style={color=black, thick},
	facet/.style={fill=white,fill opacity=0.300000},
	vertex/.style={inner sep=1pt,circle,draw=red!25!black,fill=red!75!black,thick}]
%
%
%% This TikZ-picture was produce with Sagemath version 9.3
%% with the command: ._tikz_2d_in_3d and parameters:
%% view = [-0.187100000000000, -0.451600000000000, -0.872400000000000]
%% angle = 140.260000000000
%% scale = 3
%% edge_color = black
%% facet_color = white
%% opacity = 0.300000000000000
%% vertex_color = red
%% axis = False

%% Coordinate of the vertices:
%%
\coordinate (1.00000, 0.63093, 0.50000) at (1.00000, 0.63093, 0.50000);
\coordinate (1.00000, 0.50000, 0.63093) at (1.00000, 0.50000, 0.63093);
\coordinate (0.63093, 1.00000, 0.50000) at (0.63093, 1.00000, 0.50000);
\coordinate (0.63093, 0.50000, 1.00000) at (0.63093, 0.50000, 1.00000);
\coordinate (0.50000, 1.00000, 0.63093) at (0.50000, 1.00000, 0.63093);
\coordinate (0.50000, 0.63093, 1.00000) at (0.50000, 0.63093, 1.00000);
%%
%%
%% Drawing the interior
%%
\fill[facet] (0.50000, 0.63093, 1.00000) -- (0.63093, 0.50000, 1.00000) -- (1.00000, 0.50000, 0.63093) -- (1.00000, 0.63093, 0.50000) -- (0.63093, 1.00000, 0.50000) -- (0.50000, 1.00000, 0.63093) -- cycle {};
%%
%%
%% Drawing edges
%%
\draw[edge] (1.00000, 0.63093, 0.50000) -- (1.00000, 0.50000, 0.63093);
\draw[edge] (1.00000, 0.63093, 0.50000) -- (0.63093, 1.00000, 0.50000);
\draw[edge] (1.00000, 0.50000, 0.63093) -- (0.63093, 0.50000, 1.00000);
\draw[edge] (0.63093, 1.00000, 0.50000) -- (0.50000, 1.00000, 0.63093);
\draw[edge] (0.63093, 0.50000, 1.00000) -- (0.50000, 0.63093, 1.00000);
\draw[edge] (0.50000, 1.00000, 0.63093) -- (0.50000, 0.63093, 1.00000);
%%
%%
%% Drawing the vertices
%%
\node[vertex] at (1.00000, 0.63093, 0.50000)     {};
\node[vertex] at (1.00000, 0.50000, 0.63093)     {};
\node[vertex] at (0.63093, 1.00000, 0.50000)     {};
\node[vertex] at (0.63093, 0.50000, 1.00000)     {};
\node[vertex] at (0.50000, 1.00000, 0.63093)     {};
\node[vertex] at (0.50000, 0.63093, 1.00000)     {};
\end{tikzpicture}
        \caption{$\ndoc=3$. A polygon in $\R^3$}
    \end{subfigure}
    \hfill
    \begin{subfigure}[b]{0.3\textwidth}
        \centering
        \begin{tikzpicture}%
	[x={(-0.814949cm, -0.496765cm)},
	y={(0.579533cm, -0.698504cm)},
	z={(-0.000061cm, 0.515089cm)},
	scale=5.000000,
	back/.style={loosely dotted, thin},
	edge/.style={color=black, thick},
	facet/.style={fill=white,fill opacity=0.300000},
	vertex/.style={inner sep=1pt,circle,draw=red!25!black,fill=red!75!black,thick}]
%
%
%% This TikZ-picture was produce with Sagemath version 9.3
%% with the command: ._tikz_3d_in_3d and parameters:
%% view = [-0.0850000000000000, -0.266300000000000, -0.960100000000000]
%% angle = 145.910000000000
%% scale = 3
%% edge_color = black
%% facet_color = white
%% opacity = 0.300000000000000
%% vertex_color = red
%% axis = False

%% Coordinate of the vertices:
%%
\coordinate (-0.01094, -0.17582, -0.40257) at (-0.01094, -0.17582, -0.40257);
\coordinate (-0.01094, -0.26073, -0.35355) at (-0.01094, -0.26073, -0.35355);
\coordinate (-0.16212, -0.06892, -0.40257) at (-0.16212, -0.06892, -0.40257);
\coordinate (-0.16212, -0.31418, -0.26097) at (-0.16212, -0.31418, -0.26097);
\coordinate (-0.24217, -0.09722, -0.35355) at (-0.24217, -0.09722, -0.35355);
\coordinate (-0.24217, -0.25758, -0.26097) at (-0.24217, -0.25758, -0.26097);
\coordinate (0.41523, -0.02515, -0.14160) at (0.41523, -0.02515, -0.14160);
\coordinate (0.41523, -0.11005, -0.09258) at (0.41523, -0.11005, -0.09258);
\coordinate (-0.16212, 0.38310, -0.14160) at (-0.16212, 0.38310, -0.14160);
\coordinate (-0.16212, -0.31418, 0.26097) at (-0.16212, -0.31418, 0.26097);
\coordinate (-0.24217, 0.35480, -0.09258) at (-0.24217, 0.35480, -0.09258);
\coordinate (-0.24217, -0.25758, 0.26097) at (-0.24217, -0.25758, 0.26097);
\coordinate (0.41523, 0.13520, -0.04902) at (0.41523, 0.13520, -0.04902);
\coordinate (0.41523, -0.11005, 0.09258) at (0.41523, -0.11005, 0.09258);
\coordinate (-0.01094, 0.43655, -0.04902) at (-0.01094, 0.43655, -0.04902);
\coordinate (-0.01094, -0.26073, 0.35355) at (-0.01094, -0.26073, 0.35355);
\coordinate (-0.24217, 0.35480, 0.09258) at (-0.24217, 0.35480, 0.09258);
\coordinate (-0.24217, -0.09722, 0.35355) at (-0.24217, -0.09722, 0.35355);
\coordinate (0.41523, 0.13520, 0.04902) at (0.41523, 0.13520, 0.04902);
\coordinate (0.41523, -0.02515, 0.14160) at (0.41523, -0.02515, 0.14160);
\coordinate (-0.01094, 0.43655, 0.04902) at (-0.01094, 0.43655, 0.04902);
\coordinate (-0.01094, -0.17582, 0.40257) at (-0.01094, -0.17582, 0.40257);
\coordinate (-0.16212, 0.38310, 0.14160) at (-0.16212, 0.38310, 0.14160);
\coordinate (-0.16212, -0.06892, 0.40257) at (-0.16212, -0.06892, 0.40257);
%%
%%
%% Drawing edges in the back
%%
\draw[edge,back] (-0.01094, -0.17582, -0.40257) -- (-0.01094, -0.26073, -0.35355);
\draw[edge,back] (-0.01094, -0.17582, -0.40257) -- (-0.16212, -0.06892, -0.40257);
\draw[edge,back] (-0.01094, -0.17582, -0.40257) -- (0.41523, -0.02515, -0.14160);
\draw[edge,back] (-0.01094, -0.26073, -0.35355) -- (-0.16212, -0.31418, -0.26097);
\draw[edge,back] (-0.01094, -0.26073, -0.35355) -- (0.41523, -0.11005, -0.09258);
\draw[edge,back] (-0.16212, -0.06892, -0.40257) -- (-0.24217, -0.09722, -0.35355);
\draw[edge,back] (-0.16212, -0.06892, -0.40257) -- (-0.16212, 0.38310, -0.14160);
\draw[edge,back] (-0.16212, -0.31418, -0.26097) -- (-0.24217, -0.25758, -0.26097);
\draw[edge,back] (-0.16212, -0.31418, -0.26097) -- (-0.16212, -0.31418, 0.26097);
\draw[edge,back] (-0.24217, -0.09722, -0.35355) -- (-0.24217, -0.25758, -0.26097);
\draw[edge,back] (-0.24217, -0.09722, -0.35355) -- (-0.24217, 0.35480, -0.09258);
\draw[edge,back] (-0.24217, -0.25758, -0.26097) -- (-0.24217, -0.25758, 0.26097);
%%
%%
%% Drawing vertices in the back
%%
\node[vertex] at (-0.01094, -0.17582, -0.40257)     {};
\node[vertex] at (-0.16212, -0.06892, -0.40257)     {};
\node[vertex] at (-0.01094, -0.26073, -0.35355)     {};
\node[vertex] at (-0.16212, -0.31418, -0.26097)     {};
\node[vertex] at (-0.24217, -0.09722, -0.35355)     {};
\node[vertex] at (-0.24217, -0.25758, -0.26097)     {};
%%
%%
%% Drawing the facets
%%
\fill[facet] (-0.16212, -0.06892, 0.40257) -- (-0.24217, -0.09722, 0.35355) -- (-0.24217, 0.35480, 0.09258) -- (-0.16212, 0.38310, 0.14160) -- cycle {};
\fill[facet] (-0.16212, -0.06892, 0.40257) -- (-0.24217, -0.09722, 0.35355) -- (-0.24217, -0.25758, 0.26097) -- (-0.16212, -0.31418, 0.26097) -- (-0.01094, -0.26073, 0.35355) -- (-0.01094, -0.17582, 0.40257) -- cycle {};
\fill[facet] (-0.16212, 0.38310, 0.14160) -- (-0.24217, 0.35480, 0.09258) -- (-0.24217, 0.35480, -0.09258) -- (-0.16212, 0.38310, -0.14160) -- (-0.01094, 0.43655, -0.04902) -- (-0.01094, 0.43655, 0.04902) -- cycle {};
\fill[facet] (-0.01094, -0.17582, 0.40257) -- (-0.01094, -0.26073, 0.35355) -- (0.41523, -0.11005, 0.09258) -- (0.41523, -0.02515, 0.14160) -- cycle {};
\fill[facet] (-0.16212, -0.06892, 0.40257) -- (-0.01094, -0.17582, 0.40257) -- (0.41523, -0.02515, 0.14160) -- (0.41523, 0.13520, 0.04902) -- (-0.01094, 0.43655, 0.04902) -- (-0.16212, 0.38310, 0.14160) -- cycle {};
\fill[facet] (-0.01094, 0.43655, 0.04902) -- (-0.01094, 0.43655, -0.04902) -- (0.41523, 0.13520, -0.04902) -- (0.41523, 0.13520, 0.04902) -- cycle {};
\fill[facet] (0.41523, -0.02515, 0.14160) -- (0.41523, -0.11005, 0.09258) -- (0.41523, -0.11005, -0.09258) -- (0.41523, -0.02515, -0.14160) -- (0.41523, 0.13520, -0.04902) -- (0.41523, 0.13520, 0.04902) -- cycle {};
%%
%%
%% Drawing edges in the front
%%
\draw[edge] (0.41523, -0.02515, -0.14160) -- (0.41523, -0.11005, -0.09258);
\draw[edge] (0.41523, -0.02515, -0.14160) -- (0.41523, 0.13520, -0.04902);
\draw[edge] (0.41523, -0.11005, -0.09258) -- (0.41523, -0.11005, 0.09258);
\draw[edge] (-0.16212, 0.38310, -0.14160) -- (-0.24217, 0.35480, -0.09258);
\draw[edge] (-0.16212, 0.38310, -0.14160) -- (-0.01094, 0.43655, -0.04902);
\draw[edge] (-0.16212, -0.31418, 0.26097) -- (-0.24217, -0.25758, 0.26097);
\draw[edge] (-0.16212, -0.31418, 0.26097) -- (-0.01094, -0.26073, 0.35355);
\draw[edge] (-0.24217, 0.35480, -0.09258) -- (-0.24217, 0.35480, 0.09258);
\draw[edge] (-0.24217, -0.25758, 0.26097) -- (-0.24217, -0.09722, 0.35355);
\draw[edge] (0.41523, 0.13520, -0.04902) -- (-0.01094, 0.43655, -0.04902);
\draw[edge] (0.41523, 0.13520, -0.04902) -- (0.41523, 0.13520, 0.04902);
\draw[edge] (0.41523, -0.11005, 0.09258) -- (-0.01094, -0.26073, 0.35355);
\draw[edge] (0.41523, -0.11005, 0.09258) -- (0.41523, -0.02515, 0.14160);
\draw[edge] (-0.01094, 0.43655, -0.04902) -- (-0.01094, 0.43655, 0.04902);
\draw[edge] (-0.01094, -0.26073, 0.35355) -- (-0.01094, -0.17582, 0.40257);
\draw[edge] (-0.24217, 0.35480, 0.09258) -- (-0.24217, -0.09722, 0.35355);
\draw[edge] (-0.24217, 0.35480, 0.09258) -- (-0.16212, 0.38310, 0.14160);
\draw[edge] (-0.24217, -0.09722, 0.35355) -- (-0.16212, -0.06892, 0.40257);
\draw[edge] (0.41523, 0.13520, 0.04902) -- (0.41523, -0.02515, 0.14160);
\draw[edge] (0.41523, 0.13520, 0.04902) -- (-0.01094, 0.43655, 0.04902);
\draw[edge] (0.41523, -0.02515, 0.14160) -- (-0.01094, -0.17582, 0.40257);
\draw[edge] (-0.01094, 0.43655, 0.04902) -- (-0.16212, 0.38310, 0.14160);
\draw[edge] (-0.01094, -0.17582, 0.40257) -- (-0.16212, -0.06892, 0.40257);
\draw[edge] (-0.16212, 0.38310, 0.14160) -- (-0.16212, -0.06892, 0.40257);
%%
%%
%% Drawing the vertices in the front
%%
\node[vertex] at (0.41523, -0.02515, -0.14160)     {};
\node[vertex] at (0.41523, -0.11005, -0.09258)     {};
\node[vertex] at (-0.16212, 0.38310, -0.14160)     {};
\node[vertex] at (-0.16212, -0.31418, 0.26097)     {};
\node[vertex] at (-0.24217, 0.35480, -0.09258)     {};
\node[vertex] at (-0.24217, -0.25758, 0.26097)     {};
\node[vertex] at (0.41523, 0.13520, -0.04902)     {};
\node[vertex] at (0.41523, -0.11005, 0.09258)     {};
\node[vertex] at (-0.01094, 0.43655, -0.04902)     {};
\node[vertex] at (-0.01094, -0.26073, 0.35355)     {};
\node[vertex] at (-0.24217, 0.35480, 0.09258)     {};
\node[vertex] at (-0.24217, -0.09722, 0.35355)     {};
\node[vertex] at (0.41523, 0.13520, 0.04902)     {};
\node[vertex] at (0.41523, -0.02515, 0.14160)     {};
\node[vertex] at (-0.01094, 0.43655, 0.04902)     {};
\node[vertex] at (-0.01094, -0.17582, 0.40257)     {};
\node[vertex] at (-0.16212, 0.38310, 0.14160)     {};
\node[vertex] at (-0.16212, -0.06892, 0.40257)     {};
\end{tikzpicture}
        \caption{$\ndoc=4$. A polyhedron in $\R^4$}
    \end{subfigure}
    \caption{Examples of the \acrshort{dcg} expohedron for $\ndoc\in\{2, 3, 4\}$ items. The vertices are the \acrshort{dcg} exposures $\left(\frac{1}{\logtwo(2)},\hdots,\frac{1}{\logtwo(\ndoc+1)}\right)$ under application of the symmetric group $\cS_\ndoc$.
    The expohedron is the convex hull of these vertices.
    Expohedra live in hyperplanes of dimension $\ndoc-1$.}
    \Description{Expohedron illustrations}
    \label{fig:expohedron}
\end{teaserfigure}
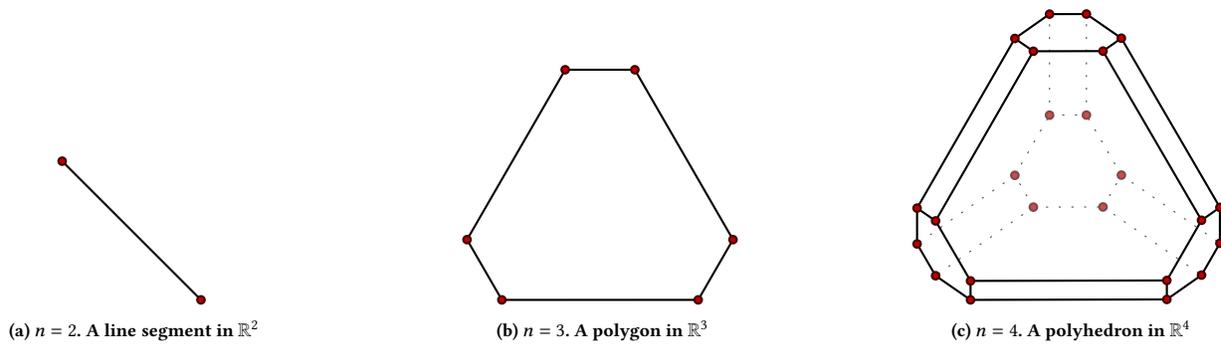

\maketitle

\section{Introduction}
    Ranking systems are nowadays widely deployed in the \acrlong{www} and make it possible for users of search engines to quickly retrieve items that are relevant \acrlong{wrt} their query.
Those items can be websites, movies, songs, research papers and many other things.
With web search and recommendation having an ever increasing influence on people's behavior and determining which opportunities people are given, the questions whether ranking systems are \emph{fair} and how to make them fair, has gathered much attention in recent years \cite{singh_fairness_2018,biega_equity_2018,pitoura_fairness_2021}.
% A detailed description of where our contribution fits in the large repertoire of fairness definition and methods, is available in section \ref{sec:related_work}, where we discuss related work.

% What exactly it means for a system to be fair is a question as old as justice itself, on which philosophers have pondered for centuries, and to which we have no ambition to present an answer.
% Rather we provide a mathematical tool that is useful for some potential notions of fairness.

A typical ranking system estimates a relevance value for each item and orders the items by decreasing relevance values; this is called the \acrfull{prp} and provides under some assumptions the highest expected utility for the users \cite{robertson_probability_1977}.
However, it has been noted that \acrshort{prp} rankings can be very unfair to item producers \cite{singh_fairness_2018,biega_equity_2018} insofar as their exposure is not proportional to their merit.
For instance when all items are almost equally relevant, the last item will receive much less exposure than the first item, even though their merits are almost equal.
To avoid this unfairness, a common approach is to deliver different rankings each time a user issues the query, in such a way that the average exposure given to each item is proportional to its merit.
This is called \emph{amortization} \cite{biega_equity_2018}.
% With amortization it is possible to reduce the unfairness, but at a price of reduced utility.
% The natural question at this stage is to ask:

% \emph{Which are the proportions with which each possible ranking should be delivered?}
% \emph{Which is the best distribution over rankings that the ranking system should deliver ?}

% In recent work \citeauthor{singh_fairness_2018} provide an answer to this question in \cite{singh_fairness_2018}.
In recent work, Singh and Joachims give a method to compute a distribution over rankings that maximizes utility while satisfying a fairness constraint \cite{singh_fairness_2018}.
They express a distribution over rankings as a bistochastic matrix whose elements represent the proportions with which an item should be at a certain rank.
They formulate the utility objective as a linear function of a bistochastic matrix and the fairness objective as a linear constraint on the bistochastic matrix.
% This \acrfull{lp} is solved and the solution corresponds to a Pareto-optimal point with minimal unfairness.
The optimal bistochastic matrix is then decomposed as a distribution over permutations using the \acrfull{bvn} algorithm \cite{dufosse_notes_2016}.
A similar approach has been used in follow-up work \cite{wang_user_2021,su_optimizing_2021,singh_fairness_2021}.
However it has been noted that such an approach does not scale well \cite{geyik_fairness-aware_2019} as it involves a linear program (LP) over $\ndoc^2$ variables, where $\ndoc$ is the number of items to rank.
Using state-of-the-art \acrshort{lp} solvers adapted to our problem, the total complexity of such an approach is then $O(\ndoc^5)$.
% As state-of-the-art LP solvers have $O(d^{2.5})$ complexity with $d$ the number of decision variables and assuming that the number of constraints is in $O(d)$, determining the solution therefore has complexity at least $O((\ndoc^2)^{5/2})=O(\ndoc^5)$. The question remains whether it is possible to compute optimal distributions over permutations in a more efficient way.

In this paper we lay out a method that computes the whole Pareto-frontier of the fairness-utility problem, in less time than it takes an \acrshort{lp} to compute just one endpoint of the frontier, both in terms of algorithmic complexity and empirical runtime.
Furthermore, our method is applicable to any reasonable notion of merit, requiring only that more relevant items have greater or equal merit than less relevant items.
% Our solution being independent of any particular notion of individual fairness, it could potentially be applied to any
% In this paper we give an answer to this questions under certain assumptions.

Our solution strongly relies on the introduction of a geometrical object, a polytope that we call \emph{expohedron}, as it generalizes the well-known permutohedron.
While the permutohedron is defined as the convex hull of all permutations of the vector $(1,\hdots,\ndoc)\in\R^\ndoc$, the $\vgamma$-expohedron is defined as the convex hull of all permutations of an arbitrary vector $\vgamma\in\R^\ndoc$.
Figure \ref{fig:expohedron} illustrates some expohedra for $\ndoc\in\{2,3,4\}$.
We are not the first to study this kind of polytope, also called \emph{orbit polytope} \cite{onn_geometry_1993}, but to the best of our knowledge, we are the first to apply it to a fair ranking problem.
When we set $\vgamma$ to be the vector of exposures \acrshort{aka} examination parameters of a \acrfull{pbm}, each vertex of the $\vgamma$-expohedron represents the exposures allocated to the items by a certain ranking.
A vector obtained by a convex combination of $N$ vertices represents the average exposure obtained by the items when delivering each of the $N$ rankings with proportions given by the convex coefficients.

% Our contribution is threefold:
% Based on the expohedron, we make the following main technical contributions:
Our main technical contributions are:
\begin{enumerate}
    \item\label{enum:contr_gls} We propose an efficient algorithm that takes any target exposure inside the expohedron as input and finds a distribution over $\ndoc$ rankings whose expected value is equal to the target exposure, where $\ndoc$ is the number of items to rank.
    Such an algorithm is called a \emph{Carathéodory decomposition} \cite{caratheodory_uber_1907,naszodi_perron_2019}.
    Our algorithm has complexity $O(\ndoc^2\log(\ndoc))$ and is to the best of our knowledge the first Carathéodory decomposition algorithm in the expohedron.
    \item\label{enum:contr_pareto} We propose an efficient algorithm that finds all Pareto-optimal exposure vectors in the expohedron, given a linear utility objective and a quadratic fairness objective as in \cite{diaz_evaluating_2020}, relying only on simple geometrical constructions.
    %We give an algorithm able to analytically solve the multi-objective utility-fairness optimization problem to complete Pareto-optimality.
    The result of this algorithm is a set of exposure vectors, each of which can be expressed as the expectation of a distribution using contribution (\ref{enum:contr_gls}).
    This algorithm has complexity $O(\ndoc^2\log(n))$.%, with $\ndoc$ being the number of items to rank.
    \item\label{enum:contr_billiard} As an alternative to randomly \emph{sampling} rankings from a distribution, we propose to use a deterministic scheduling method based on balanced words \cite{vuillon_balanced_2003}.
    Such an approach, as we will argue, better approximates the actual proportions in the distribution, and thus decreases the variance around the expected value when the rankings are actually delivered.
    \item We perform experiments on synthetic data and on the \acrshort{trec}2020 Fairness Track \cite{trec-fair-ranking-2019} and \acrshort{mslr} \cite{DBLP:journals/corr/QinL13} datasets to compare our methods to several baselines in terms of Pareto-optimality and efficiency.
\end{enumerate}

% The remainder of this paper is structured as follows:
% In Section \ref{sec:related_work} we discuss related work.
% In Section \ref{sec:problem} we formalize the problem and rigorously define our two objectives: user-utility and item-fairness.
% In Section \ref{sec:expohedron} we formally introduce the expohedron, we lay out some of its properties in \ref{sec:combinatorial_structure}, and present a Carathéodory decomposition algorithm in \ref{sec:caratheodory}.
% In section \ref{sec:pareto} we explain how we can use the expohedron to find the Pareto-curve.
% In Section \ref{sec:billiard} we show how we can deploy a policy, once a point on the Pareto-curve is chosen.
% In section \ref{sec:experiments}, we present some experiments we conducted on synthetic data and on the \acrshort{trec} and \acrshort{mslr} datasets.
% Finally in section \ref{sec:conclusion} we draw conclusions and propose some perspectives for future work.

\section{Related work}\label{sec:related_work}
    \paragraph{Fairness in ranking} There is a large body of literature that studies problems of fairness in \acrshort{ai} systems.
Several applications are studied, such as classification \cite{dwork_fairness_2012} or regression \cite{narasimhan_pairwise_2020}.
In the context of ranking, there are pre-processing methods \cite{zehlike_reducing_2020,salimi_interventional_2019}, whose aim is to "de-bias" the data before it is processed by a ranking algorithm; in-processing methods \cite{zehlike_reducing_2020}, that aim to modify the ranking algorithm itself; and post-processing methods, which re-order the items only in the end \cite{singh_fairness_2018,biega_equity_2018,zehlike_fa*ir:_2017}.
Our work is a post-processing approach.

Fairness can be enforced on the side of the ranking recipients \cite{do_online_2021} or on the side of the item producers \cite{singh_fairness_2018,biega_equity_2018,zehlike_fa*ir:_2017}.
We consider fairness on the item producer side.

Fairness of exposure can be enforced at the level of groups \cite{singh_fairness_2018} with the idea that each group of items should receive a certain amount of exposure, or at the level of the individual with the idea that every item \cite{diaz_evaluating_2020} or every item producer \cite{biega_equity_2018} should receive a certain amount of exposure.
This is called \emph{individual fairness} \cite{biega_equity_2018}.
In this paper we use individual fairness at the level of the items, \acrshort{ie} we ensure that each item gets a certain target exposure.

There are many published definitions of fairness.
Some \emph{group fairness} definitions \cite{zehlike_fa*ir:_2017} look at the items in the top-$k$ ranks and enforce that each group is present with equal proportions for every $k$.
Other approaches called \emph{fairness of exposure} \cite{biega_equity_2018,wang_fairness_2021,oosterhuis_computationally_2021,singh_fairness_2018,morik_controlling_2020,su_optimizing_2021,singh_policy_2019,diaz_evaluating_2020,wu_tfrom_2021,wang_user_2021}, like our own approach, seek to adjust the exposure of the items, to a certain range of admissible or optimal values.
This is the case with a \acrlong{ltr} approach by Oosterhuis \cite{oosterhuis_computationally_2021} that applies gradient-descent to find the optimal scores of the items when the ranking policy is a \acrlong{pl} distribution.

Approaches using fairness of exposure depend on the exposure model that is used.
In a \acrfull{pbm}, the exposure of an item depends only on its rank \cite{chuklin_click_2015}.
% With other models such as \acrfull{dbn} \cite{chapelle_dynamic_2009} or \acrfull{ubm} \cite{dupret_user_2008}, the exposure of an item also depends on higher ranked items.
With other models such as \acrfull{dbn} \cite{chapelle_dynamic_2009,chuklin_click_2015}, the exposure of an item also depends on higher ranked items.
Our approach is applicable to \acrshort{pbm}s.

\paragraph{Optimizing bistochastic matrices}
Singh and Joachims \cite{singh_fairness_2018} formulate a ranking policy as a bistochastic matrix whose elements are the probabilities for an item to end up at a certain rank.
When a \acrshort{pbm} is used, it is possible to formulate fairness as a linear constraint and utility as a linear objective.
The problem is solved using a linear program (\acrshort{lp}) and an optimal bistochastic matrix is found.
It is then possible to decompose the bistochastic matrix as a convex sum of permutation matrices.
This is called a \acrfull{bvn} decomposition and expresses the bistochastic matrix as an expected value of a distribution over permutations.
The ranking policy then consists in sampling from this distribution.
This approach has later been adopted in subsequent papers \cite{wang_user_2021,su_optimizing_2021,singh_fairness_2021}.
Such an approach finds only one point of the Pareto-frontier: the point with minimal unfairness, whereas our approach takes less time to find the whole frontier.
This is achieved by working in the expohedron instead of the Birkhoff polytope (\acrshort{ie} the set of bistochastic matrices).

\paragraph{Controller}
Another more heuristic approach is to use controllers.
A controller generally starts by delivering a \acrshort{prp} ranking and computes at every time-step the exposure difference \acrshort{wrt} a certain target value.
This error term is added to the relevance scores and for the next ranking the items are ordered by this new score.
In other words, a controller greedily corrects the empirical unfairness at every time-step by giving adequate bonuses to disadvantaged items.
This can be done in a static setting \cite{thonet_multi-grouping_2020} or within a \acrlong{ltr} framework \cite{morik_controlling_2020}.

% add controller

\section{Model and problem statement}\label{sec:problem}
    % In order to properly state our problem, we first need to introduce some concepts.

\paragraph{Setting}
We suppose that we are given a query $\query$ to which is associated a set of $\ndoc$ items indexed by $\ixitem$.
In our setting, all queries are treated independently.
Therefore in the remainder of this paper, we omit indexing with $\query$.
% We further suppose that we are given for each item $\ixitem$ a relevance value $\vrho_i\in[0,1]$ that expresses how much the item is relevant to the query.
We further suppose that we are given a vector of relevance values $\vrho\in[0,1]^\ndoc$ that expresses how much each item is relevant to the query.
We assume that the query is repeated by the users many times, giving us sufficiently many opportunities to execute our amortization.

\paragraph{Ranking}
In this paper we denote $\cS_\ndoc$ the set of all permutation matrices of size $\ndoc$.
We define a ranking as a permutation matrix $\pi\in\cS_\ndoc$ such that $\pi_{\ixitem\ixrank}=1$ if and only if the item $\ixitem$ is at rank $\ixrank$.

\paragraph{\acrshort{pbm} exposure models}
We work with a \acrfull{pbm} \cite{chuklin_click_2015}, represented by a vector $\vgamma\in\R_+^\ndoc$ whose $\ixrank^\text{th}$ component is the exposure associated to rank $\ixrank$.
We assume \acrshort{wlog}\footnote{It is \acrshort{wlog}, because we can just rename the ranks to make it true.} that the exposure decreases with increasing rank.
The exposure is a measure of how much attention the users dedicate to a certain rank.

% \paragraph{Exposure vector}
Given a ranking $\pi\in\cS_\ndoc$, the exposure associated to the items is given by the vector $\vcE(\pi) = \pi\vgamma$, where the $\ixitem^\text{th}$ element of $\vcE(\pi)$ is the exposure allocated to item $\ixitem$.

Given a distribution $\cD$ of $N$ rankings $\pi_1,\hdots,\pi_N$ delivered with proportions $p_1,\hdots,p_N$, the expected exposure associated to the items and denoted $\vcE(\cD)$ is given by
\begin{equation}
    \vcE(\cD) = \sum_{i=1}^N p_i\vcE(\pi_i) = \sum_{i=1}^N p_i\pi_i\vgamma.
\end{equation}

\paragraph{Utility}
Given a ranking $\pi\in\cS_\ndoc$, the utility is the scalar product of the relevance vector with the exposures associated to the items:

\begin{equation}
    U(\pi) \defeq \vrho^\top \vcE(\pi) = \vrho^\top \pi\vgamma.
\end{equation}
When one chooses the $\ixrank^\text{th}$ element of $\vgamma$ to be $\gamma_k = 1 / \logtwo(k+1)$ for all $k\in\seq{1}{\ndoc}$, then $U(\pi)$ is the \acrfull{dcg} metric \cite{jarvelin_cumulated_2002} of ranking $\pi$.
When one chooses $\gamma_k = (1-p)p^{k-1}$ for some $p\in(0,1)$, then $U$ is the \acrfull{rbp} metric \cite{moffat_rank-biased_2008} of ranking $\pi$.
% This definition of user utility is of course debatable, but such a discussion is not the subject of this paper.
Given a distribution $\cD$ of $N$ rankings $\pi_1,\hdots,\pi_N$ delivered with proportions $p_1,\hdots,p_N$, the average utility is
\begin{equation}
    U(\cD) = \sum_{i=1}^N p_i U(\pi) = \sum_{i=1}^N p_i\vrho^\top \pi_i\vgamma = \vrho^\top \vcE(\cD).
\end{equation}
% Note that $U(\cD)$ depends on $\cD$ only through its expectation $\vcE(\cD)$, so that we can write $U(\vcE(\cD))$.

\paragraph{(Un)fairness}
% Defining fairness is always a delicate matter.
% There is no consensus on what fairness means from a philosophical standpoint.
Since there exist multiple definitions of fairness, in this paper we assume that a decision-maker has decided which exposures are fair by determining a \emph{target exposure} for each item, for instance as in \cite{diaz_evaluating_2020} where the target exposure is defined as an affine function of the relevance vector $\vrho$ (some kind of meritocratic fairness).
This target is to be understood as the amount of exposure an item deserves each time the query is made.
% Thus our approach is applicable to any way of defining the desert of items.
If this target exposure is reached then the system is fair by definition.
In order to define how unfair a system is we need a function that measures how far we are from the target exposure.
In this paper we proceed as in \cite{diaz_evaluating_2020} in adopting the standard euclidean norm.
% The extension to $\ell^p$ norms for $p\in[1,+\infty]$ is left for future work.
Formally, if we denote $\vcEt$ the target exposure and $\vcE(\cD)$ the actual expected exposure allocated to the items, then the unfairness is
\begin{align}
    F(\cD,\vcEt) \defeq& \norm{\vcE(\cD) - \vcEt}_2.
\end{align}
% Note that the unfairness can be expressed as a function of the expected exposure $\vcE(\cD)$ of $\cD$.
% Note that the unfairness depends on $\cD$ only through its expectation $\vcE(\cD)$, so that we can write $F(\vcE(\cD),\vcEt)$.
While in \cite{singh_fairness_2018}, fairness is defined as a condition that is either satisfied or not, we consider here a way to quantify it.
This will allow us to trade off utility with fairness.

\paragraph{Objective}

Note that both $U$ anf $F$ depend on $\cD$ only through its expectation so that we can write $U(\vcE(\cD))$ and $F(\vcE(\cD), \vcEt)$.
Our goal is to solve a \acrfull{moo} problem with two objectives, the maximization of utility and the minimization of unfairness after the delivery of a large number of rankings:

\begin{align}\label{eq:moo}
\begin{split}
    \max_{\cD} U(\vcE(\cD)),~
    \min_{\cD} F(\vcE(\cD), \vcEt).
\end{split}
\end{align}

In principle one would expect the optimization variable to be a sequence of rankings.
However we  decompose the problem into 3 distinct sub-steps.
\begin{enumerate}
    \item\label{enum:moo} Find all Pareto-optimal expected exposure vectors $\vcE$.
    We solve this problem in Section \ref{sec:pareto}.
    \item\label{enum:caratheodory} Given an expected exposure vectors $\vcE$, find a probability distribution $\cD$ over $\ndoc$ rankings, whose expectation $\vcE(\cD)$ is equal to $\vcE$.
    We solve this problem in Section \ref{sec:caratheodory}.
    \item\label{enum:billiard} Deliver the rankings of the support of $\cD$ on a finite number of samples, with proportions as close as possible to the probabilities in $\cD$.
    We solve this problem in Section \ref{sec:billiard}.
\end{enumerate}

This procedure is analogous to the one used by Singh and Joachims \cite{singh_fairness_2018}. In their paper Step (\ref{enum:moo}) is done using linear programming, but only an endpoint of the Pareto-frontier is determined and it is empirically found to be slower than our method.
Step (\ref{enum:caratheodory}) is done using a \acrshort{bvn} decomposition algorithm \cite{dufosse_notes_2016}, but the decomposition is done over at most $(\ndoc-1)^2+1$ rankings, whereas we use at most $\ndoc$ and are much quicker doing so.
Step (\ref{enum:billiard}) is done using sampling, which does not very well approximate the target proportions at low time horizons.
% In fact this procedure bears much resemblance with the procedure introduced in \cite{singh_fairness_2018}, with the chief difference being that in \cite{singh_fairness_2018} optimization is done on the rankings represented as a bistochastic matrices, whereas we optimize directly on the exposure, which eliminates the need for a \acrfull{lp}.

\section{The expohedron and a Carathéodory decomposition}\label{sec:expohedron}
    
\paragraph{The permutation simplex}
A naive way of representing distributions over permutations is to consider the space $\R^{\ndoc!}$, with each basis vector representing one permutation in $\cS_\ndoc$.
Every point in the $\R^{\ndoc!}$-simplex then corresponds to exactly one distribution over permutations, given by its decomposition into basis vectors.
Furthermore the $\R^{\ndoc!}$-simplex is the convex hull of all permutations as represented in the space $\R^{\ndoc!}$.
In practice this approach is not feasible, because of the high dimensionality of the space $\R^{\ndoc!}$.

\paragraph{The Birkhoff polytope}
In the case of \acrshort{pbm} exposure models, we can work in a smaller, more tractable space: the space of bistochastic matrices \acrshort{aka} Birkhoff polytope \cite{singh_fairness_2018}.
Indeed in order to know the expected exposure an item gets from a distribution over rankings, it suffices to know the marginal probabilities to end up at each rank.
This reduces the dimensionality of the space to $\ndoc^2$.
The Birkhoff polytope is also the convex hull of all permutations, represented by permutation matrices.
There exists an algorithm, called \acrfull{bvn} decomposition \cite{dufosse_notes_2016}, that can express any bistochastic matrix of size $\ndoc\times\ndoc$ as a convex sum of at most $(\ndoc-1)^2+1$ permutation matrices.
% However this decomposition can b

\paragraph{The expohedron}
It is possible to reduce even further the dimensionality of the space we work in.
Indeed what interests us in the end is the exposure that the items get, not with which frequency they are at certain ranks.
For a \acrshort{pbm} with exposure vector $\vgamma$, the $\vgamma$-expohedron is formally defined as
\begin{equation}
    \expo(\vgamma) \defeq \Conv\left(\pi\vgamma~|~\pi\in\cS_\ndoc\right).
\end{equation}
% It is the convex hull of all exposure vectors achievable with a ranking.
It is the convex hull of exposure vectors achievable with all possible rankings.

\begin{theo}[Carathéodory \cite{caratheodory_uber_1907,naszodi_perron_2019}]
    Let $\Pi\in\R^\ndoc$ be the convex hull of a finite number of points $\vv_1,\hdots,\vv_N$.
    Then any point $\vx\in\Pi$ can be expressed as the convex sum of at most $\ndoc+1$ points $\vv_i$.
\end{theo}

Carathéodory's theorem tells us that it is possible to express any point in the expohedron as a convex sum of at most $\ndoc+1$ vertices.
Such a convex sum is called \emph{Carathéodory decomposition}.
In our case the sum can be made over at most $\ndoc$ vertices, because the expohedron lives within a hyperplane of $\R^\ndoc$.
% Such a convex sum is called \emph{Carathéodory decomposition}.
Indeed by permuting the elements of $\vgamma$, the sum of the elements of $\pi\vgamma$ remains unchanged.
Therefore the vector $\vone$ is orthogonal to the expohedron.
% In fact the \acrshort{bvn} algorithm is a Carathéodory decomposition algorithm in the Birkhoff polytope.

A Carathéodory decomposition can express any point $\vcE$ inside the expohedron as the expected value $\vcE(\cD)$ of a distribution $\cD$ over at most $\ndoc$ rankings.

% In the remainder of this section we exhibit some important properties of the expohedron and we finish by laying out an algorithm able to find a Carathéodory decomposition in the expohedron.

    \subsection{Properties of the expohedron}\label{sec:combinatorial_structure}
    % When $\vgamma = (1,\hdots,\ndoc)$, the $\vgamma$-expohedron is called permutohedron \cite{noauthor_permutohedron_2021}.
% The permutohedron is a well-studied object and it turns out that the expohedron inherits many of its properties.
% Indeed the permutohedron and the expohedron are both realizations of the same abstract polytope, so that all combinatorial properties are preserved.
% The complete combinatorial structure of the permutohedron (and thus of the expohedron) was given by \citeauthor{queyranne_structure_1993} \cite{queyranne_structure_1993}.
% So the complete face lattice of our expohedron is known.

% Concerning the geometrical properties, we will need the notion of majorization.

\paragraph{Majorization}
The concept of majorization \cite{marshall_inequalities_2011} gives us an easy-to-check criterion for verifying if a point is inside the expohedron.
We say that a vector $\va\in\R^\ndoc$ is \emph{majorized} by a vector $\vb\in\R^\ndoc$ and we write $\va\prec\vb$ if and only if
% \begin{subequations}
%     \begin{align}
%     \forall k\in\seq{1}{n},~\sum_{i=1}^k \vb_{(i)} \leq \sum_{i=1}^k \va_{(i)},\\
%     \sum_{i=1}^n \vb_{(i)} = \sum_{i=1}^n \va_{(i)},
%     \end{align}
% \end{subequations}
\begin{align}
% \forall k\in\seq{1}{n},~\sum_{i=1}^k \vb_{(i)} \leq \sum_{i=1}^k \va_{(i)},\quad\sum_{i=1}^n \vb_{(i)} = \sum_{i=1}^n \va_{(i)},
\forall k\in\seq{1}{n},~\sum_{i=1}^k \vb^\uparrow_{i} \leq \sum_{i=1}^k \va^\uparrow_{i},\quad\sum_{i=1}^n \vb^\uparrow_{i} = \sum_{i=1}^n \va^\uparrow_{i},
\end{align}
% where $\va_{(i)}$ and $\vb_{(i)}$ are the elements of $\va$ and $\vb$ ordered from the smallest to the greatest element.
where $\va^\uparrow$ and $\vb^\uparrow$ are the vectors $\va$ and $\vb$ with elements ordered from the smallest to the greatest.

Then one can show that a point $\vx$ is inside the $\vgamma$-expohedron if and only if it is majorized by $\vgamma$ \cite{rado_inequality_1952,marshall_inequalities_2011}.
Formally
\begin{equation}
    \vx\in\expo(\vgamma) \iff \vx\prec\vgamma.
\end{equation}

\paragraph{Order-preserving zone}
% We introduce the concept of \emph{order-preserving zone} or \emph{zone} as a partition of $\R^\ndoc$ made of $\ndoc!$ subsets, each corresponding to a permutation.
We introduce the concept of \emph{order-preserving zones} or \emph{zones} as $\ndoc!$ subsets of $\R^\ndoc$, each corresponding to a permutation.
We denote by $\zone(\pi)$ the zone corresponding to the set of vectors $\vx$ such that the elements of $\pi\vx$ are sorted in increasing order.
% An order-preserving zone is a subset of $\R^\ndoc$ in which all the points have their elements ordered in the same way. %, \acrshort{ie} all the points have the same result to the operation $\texttt{argsort}$.
Figures \ref{fig:gls} and \ref{fig:pareto_expo} illustrate the subdivision of an expohedron into 6 zones for $\ndoc=3$ documents.
% If the operation $\texttt{argsort}$ applied to a point $\vx$ has more than one possible outcomes (\acrshort{ie} $\vx$ has ties), then we consider the point to be a member of all zones corresponding to the possible outcomes.
If a point $\vx$ has ties, then we consider the point to be a member of all zones corresponding to the possible orderings.

% For example the points $(2,3,1),(6,20,1)$ are both in the same zone, the points $(2,3,1),(3,2,1)$ are in different zones.
% The point $(3,3,1)$ is in the same zone as $(2,3,1)$ and also in the same zone as $(3,2,1)$.
% The point $(1,1,1)$ is in all zones.

% \begin{figure}
%     \centering
%     % \includegraphics{}
%     \Description{Subdivision into zones}
%     \caption{Subdivision of the \acrshort{dcg} expohedron for $\ndoc=3$ into the 6 order-preserving zones.}
%     \label{fig:zones}
% \end{figure}

\paragraph{Face characterization}
Recall that a face of a polytope is a polytope on its boundary.
For instance vertices, edges and facets of a dice, are all faces of the dice.
We give an easy-to-check criterion for identifying the lowest-dimensional face in which a point of the expohedron lies.
A face of the expohedron is characterized by a zone and a subset of $\seq{1}{\ndoc}$, that we call \emph{splits}.

Specifically, the face characterized by zone $\zone(\pi)$ and by the splits $S$, is composed of all the points $\vx\in\Pi(\vgamma)$ satisfying the condition
\begin{equation}\label{eq:faces}
    \forall i\in S,~\sum_{k=1}^i(\pi\vx)_k = \sum_{k=1}^i\vgamma^\uparrow_{k},
\end{equation}
% where $\vgamma_{(k)}$ are the elements of $\vgamma$ ordered from the smallest to the greatest element.
where $\vgamma^\uparrow$ is the vector $\vgamma$ with elements ordered from the smallest to the greatest.
% This characterization is a property inherited from permutohedra \cite{queyranne_structure_1993}, although it is not a combinatorial property.
The dimension of a face is given by $\dim(F) = \ndoc - |S|$. %, where $S$ is the set of splits.
For a vertex, we have $S=\seq{1}{\ndoc}$ and its dimension is $0$.
For the whole polytope, we have $S=\{\ndoc\}$ and its dimension is $\ndoc-1$.
From the majorization property it follows that all faces of the expohedron have $\ndoc$ in their splits.

With this property it becomes easy to determine the faces in which a point $\vx$ of the expohedron lies.
It suffices to set $\pi$ such that $\pi\vx$ is sorted in increasing order and to check for which $i$ \eqref{eq:faces} holds.

    \subsection{Carathéodory decomposition}\label{sec:caratheodory}
    The expohedron is a convex hull of dimension $\ndoc-1$, so from Carathéodory's theorem \cite{caratheodory_uber_1907,naszodi_perron_2019} we know that any of its points can be expressed as a convex combination of at most $\ndoc$ of its vertices.
This is already a significant improvement over the approach using \acrshort{bvn} decomposition, as \acrshort{bvn} decompositions use at most $(\ndoc-1)^2 + 1$ permutation matrices \cite{marcus_diagonals_1959}.

There exist already some published Carathéodory decomposition algorithms for permutohedra \cite{hoeksma_efficient_2016,yasutake_online_2011}, but they cannot naturally be extended to expohedra.
Expohedra are not always zonotopes, which is a key property used in \cite{hoeksma_efficient_2016}, and they are generally not defined with $\vgamma=(1,\hdots,\ndoc)$, which is a key property used in \cite{yasutake_online_2011}.

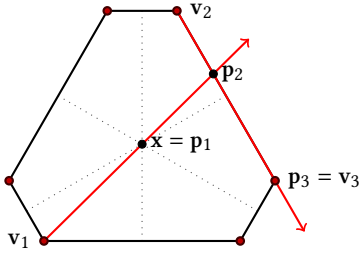
\begin{figure}
    \centering
    \begin{tikzpicture}%
	[x={(-0.707031cm, -0.408259cm)},
	y={(0.707183cm, -0.408200cm)},
	z={(0.000025cm, 0.816516cm)},
	scale=5.000000,
	back/.style={loosely dotted, thin},
	edge/.style={color=black, thick},
	facet/.style={fill=white,fill opacity=0.300000},
	vertex/.style={inner sep=1pt,circle,draw=red!25!black,fill=red!75!black,thick}]
%
%
%% This TikZ-picture was produce with Sagemath version 9.3
%% with the command: ._tikz_2d_in_3d and parameters:
%% view = [-0.187100000000000, -0.451600000000000, -0.872400000000000]
%% angle = 140.260000000000
%% scale = 3
%% edge_color = black
%% facet_color = white
%% opacity = 0.300000000000000
%% vertex_color = red
%% axis = False

%% Coordinate of the vertices:
%%
\coordinate (O) at (0., 0., 0.);
\coordinate (A) at (1.00000, 0.63093, 0.50000);
\coordinate (B) at (1.00000, 0.50000, 0.63093);
\coordinate (C) at (0.63093, 1.00000, 0.50000);
\coordinate (D) at (0.63093, 0.50000, 1.00000);
\coordinate (E) at (0.50000, 1.00000, 0.63093);
\coordinate (F) at (0.50000, 0.63093, 1.00000);
\coordinate (X) at (0.71031, 0.71031, 0.71031);
\coordinate (P) at (0.50000, 0.76794, 0.86299);
\coordinate (Pi) at (0.39484504, 0.79675283, 0.93933188);
\coordinate (Ci) at (0.5, 1.11072107, 0.52020868);

% Zone subdivision
\draw [dotted] ($(A)!(O)!(B)$) -- (O);
\draw [dotted] ($(B)!(O)!(D)$) -- (O);
\draw [dotted] ($(D)!(O)!(F)$) -- (O);
\draw [dotted] ($(F)!(O)!(E)$) -- (O);
\draw [dotted] ($(E)!(O)!(C)$) -- (O);
\draw [dotted] ($(C)!(O)!(A)$) -- (O);
%%
%%
%% Drawing the interior
%%
\fill[facet] (F) -- (D) -- (B) -- (A) -- (C) -- (E) -- cycle {};
%%
%%
%% Drawing edges
%%
\draw[edge] (A) -- (B);
\draw[edge] (A) -- (C);
\draw[edge] (B) -- (D);
\draw[edge] (C) -- (E);
\draw[edge] (D) -- (F);
\draw[edge] (E) -- (F);

\draw[->,red, thick] (A) -- (Pi);
\draw[->,red, thick] (F) -- (Ci);
\filldraw [black] (X) circle (0.3pt) node[right, black] {$\vx=\vp_1$};
\filldraw [black] (P) circle (0.3pt) node[right, black] {$\vp_2$};
%%
%%
%% Drawing the vertices
%%
\node[vertex, label=left:$\vv_1$] at (A) {};
\node[vertex] at (B)     {};
\node[vertex] at (C)     {};
\node[vertex] at (D)     {};
\node[vertex, label=right:{$\vp_3=\vv_3$}] at (E) {};
\node[vertex, label=right:$\vv_2$] at (F) {};

\end{tikzpicture}
    \caption{\acrshort{gls} procedure, inspired from an illustration in \cite{hoeksma_efficient_2016}.
    The barycenter $\vx$ of the expohedron is decomposed into a convex sum of vertices $\vv_1,\vv_2,\vv_3$.
    The dotted grey lines materialize the subdivision of the expohedron into order-preserving zones.}
    \label{fig:gls}
\end{figure}

In this section we propose to adapt a generic Carathéodory algorithm, the \acrshort{gls} method named after \acrlong{gls} \cite{grotschel_geometric_1993}.
Its main idea is illustrated in Figure \ref{fig:gls}.
% inspired from a similar figure in \cite{hoeksma_efficient_2016}.

The \acrshort{gls} method for decomposing a point $\vx$ consists in choosing an arbitrary vertex $\vv_1$ and drawing a half-line starting from $\vv_1$ and passing through $\vx$.
This half-line intersects the polytope's boundary at a point $\vp_2$ lying on a face of dimension $\ndoc-2$.
It is easy to see that $\vx$ can be expressed as a convex sum of $\vv_1$ (a vertex) and $\vp_2$ (not necessarily a vertex).
The point $\vp_2$ is then decomposed using the same method: we choose an arbitrary vertex $\vv_2$ from the face of $\vp_2$ and draw a half-line from $\vv_2$ through $\vp_2$ intersecting the polytope's boundary at a point $\vp_3$ lying on a face of dimension $\ndoc-3$ and we express $\vp_2$ as a convex sum of $\vv_2$ and $\vp_3$.
Repeating this procedure recursively, there will come a point at which the intersection $\vp_{\ndoc}$ is itself a vertex of the polytope and we are finished.

This procedure has two delicate operations that need to be adapted to the particular polytope on which it is applied:
\begin{enumerate}
    \item\label{enum:intersection} We need a method to find the intersection of a half-line with the polytope's boundary;
    \item\label{enum:face_id} For a point on a face, we need to find a vertex on the same face.
\end{enumerate}

% For the expohedron, the Step (\ref{enum:face_id}) is solved in Section \ref{sec:combinatorial_structure} in \eqref{eq:faces}.
To find a vertex on the same face as a point $\vx$, it suffices to permute the elements of $\vgamma$ such that its result to \argsort~ is the same as for $\vx$.
% To find a vertex on the same face as a point $\vx$, it suffices to choose the permutation $\pi$ such that $\vx\in\zone(\pi)$.
% The vertex then is $\pi^{-1}\vgamma_{(i)}$
This operation has complexity $O(\ndoc\log(\ndoc))$, because it is as complex as a sorting operation, and can be expressed in python\footnote{Assuming $\vgamma$ has decreasing components.} as $\vgamma[\argsort(\argsort(-\vx))]$.

% For Step (\ref{enum:intersection}), as long as we stay in an order-preserving zone, we can compute the solution analytically.
% For example we stay in an order preserving zone when we start from $\vv$ and go into a direction that has the same ordering as $\vv$.
% Let $\vv$ and $\vx$ be two points in the expohedron.
% The intersection $\vv + \lambda(\vx-\vv)$ of the half-line $[\vv\vx)$ with the expohedron must satisfy
% \begin{equation}
%     \max_{\lambda\geq1}\vv + \lambda(\vx-\vv) ~\text{s.t.}~ \vv + \lambda(\vx-\vv)\prec \vgamma.
% \end{equation}
% Through some algebraic manipulations it follows that
% \begin{equation}\label{eq:intersection}
%     \lambda = \min_k \left\{\frac{G_k - V_k}{D_k} | D_k < 0\right\},
% \end{equation}
% where $G_k = \sum_{i=1}^k \vgamma_{(i)}$, $V_k = \sum_{i=1}^k \vv_{(i)}$, $D_k = \sum_{i=1}^k (\vx-\vv)_{(i)}$, and the index $(i)$ again denotes the elements of a vector in increasing order.

For Step (\ref{enum:intersection}), one can use the majorization criterion to make a bisection search on the half-line.
Such a bisection search has complexity $O(\ndoc\log(\ndoc))$, because the sort operation to check the majorization needs to be done $O(1)$ times.

\begin{algorithm}
	\caption{The \acrshort{gls} method in the expohedron.}\label{alg:gls}
	\begin{algorithmic}[1]
		\Procedure{\texttt{GLS}}{\textsc{Input}: $\vx\in\Pi(\vgamma)$}
		  %  \State $\vv_1,\hdots,\vv_\ndoc \gets \vzero$ \Comment{Initialize the vertices of the decomposition to the null vector}
		  %  \State $\alpha_1,\hdots,\alpha_\ndoc \gets 0$ \Comment{Initialize the convex coefficients to $0$}
		  %  \State $F = $
		    \State\label{l:initial_vertex} $\vv_1\gets \vgamma[\argsort(\argsort(-\vx))]$ \Comment{Choose an initial vertex}
		    \State $\alpha_1 \gets 1$ \Comment{Set the initial vertex's weight to 1}
		    \State $\vp_1 \gets \vx$
		    \For{$i\in\seq{1}{\ndoc}$}
    % 			\State\label{l:intersection} $\lambda \gets \text{bisection_search}(\max_{\lambda\geq1}\{\vv_1+\lambda(\vx-\vv_1)\prec\vgamma\})$ \Comment{Find the intersection with the polytope's boundary}.
    			\State $\vp_{i+1} \gets \max_{\lambda\geq1}\vv_i+\lambda(\vp_{i+1}-\vv_i) \text{~s.t.~} \vv_i+\lambda(\vp_{i+1}-\vv_i)\prec\vgamma$
    			\State $\alpha_{i+1} \gets \alpha_i - \frac{\norm{\vp_i-\vp_{i+1}}}{\norm{\vp_i - \vv_i}} \alpha_i$ \Comment{Update convex coefficients}
                \State $\alpha_i \gets \frac{\norm{\vp_i-\vp_{i+1}}}{\norm{\vp_i - \vv_i}} \alpha_i$
                \State $\vv_{i+1} \gets \vgamma[\argsort(\argsort(-\vp_{i+1}))]$
    		\EndFor
		\EndProcedure~\textsc{Output}: $\alpha_1,\hdots,\alpha_\ndoc,\vv_1,\hdots,\vv_\ndoc$
	\end{algorithmic}
\end{algorithm}

\begin{theo}
    Algorithm \ref{alg:gls} find a Carathéodory decomposition of any point in the expohedron in $O(\ndoc^2\log(n))$ time.
\end{theo}
Indeed the \acrshort{gls} procedure consists in executing at most $\ndoc$ times the steps (\ref{enum:intersection}) and (\ref{enum:face_id}), which have complexity $O(\ndoc\log(\ndoc))$, so the complexity of the algorithm we propose is $O(\ndoc^2\log(n))$.
% This compares favorably to the complexity of a \acrshort{bvn} decomposition, which has complexity ???.

%\section{Our solution}\label{sec:solution}
    % \input{sections/solution}
\section{Finding the Pareto curve}\label{sec:pareto}
    In this section we show how we can solve the \acrshort{moo} problem \eqref{eq:moo}.
% We need one assumption for our method to work, namely the target exposure $\vcEt$ must be in the same zone as the \acrshort{prp} ranking.
We assume that the target exposure $\vcEt$ is in the same zone as the \acrshort{prp} ranking.
This is a weak assumption, because it is another way of saying that items with higher relevance must not get less exposure than items with lower relevance.
Furthermore we assume that the target exposure is inside the expohedron, \acrshort{ie} the target is feasible.

The decision variable $\vcE$ also needs to be inside the expohedron, in order to be the expected value of a distribution over rankings (\acrshort{ie} a convex sum of vertices).
So our constraint can be written as $\vcE \prec \vgamma$.
Formally our optimization problem is:
% \begin{subequations}\label{eq:pareto_moo}
% \begin{align}
%     \max_\vcE~& \vcE^\top \vrho,\label{eq:max}\\
%     \min_\vcE~& \norm{\vcE - \vcEt}_2,\label{eq:min}\\
%     \text{s.t.}~&\vcE\prec\vgamma.
% \end{align}
% \end{subequations}
\begin{align}\label{eq:pareto_moo}
    \max_\vcE \vrho^\top \vcE, \quad\min_\vcE \norm{\vcE - \vcEt}^2_2, \quad\text{s.t.}~\vcE\prec\vgamma.
\end{align}

The maximization objective is linear and the minimization objective is quadratic and isotropic.
The level sets of the utility objective are hyperplanes with orthogonal vector $\vrho$ and the level sets of the fairness objective are hyperspheres centered at $\vcEt$.

\begin{figure}
    \centering
    \begin{tikzpicture}%
	[x={(-0.707031cm, -0.408259cm)},
	y={(0.707183cm, -0.408200cm)},
	z={(0.000025cm, 0.816516cm)},
	scale=5.000000,
	back/.style={loosely dotted, thin},
	edge/.style={color=black, thick},
	facet/.style={fill=white,fill opacity=0.300000},
	vertex/.style={inner sep=1pt,circle,draw=red!25!black,fill=red!75!black,thick},
	extended line/.style={shorten >=-#1,shorten <=-#1},
  extended line/.default=2cm]
%
%
%% This TikZ-picture was produce with Sagemath version 9.3
%% with the command: ._tikz_2d_in_3d and parameters:
%% view = [-0.187100000000000, -0.451600000000000, -0.872400000000000]
%% angle = 140.260000000000
%% scale = 3
%% edge_color = black
%% facet_color = white
%% opacity = 0.300000000000000
%% vertex_color = red
%% axis = False

%% Coordinate of the vertices:
%%
\coordinate (O) at (0., 0., 0.);
\coordinate (A) at (1.00000, 0.63093, 0.50000);
\coordinate (B) at (1.00000, 0.50000, 0.63093);
\coordinate (C) at (0.63093, 1.00000, 0.50000);
\coordinate (D) at (0.63093, 0.50000, 1.00000);
\coordinate (E) at (0.50000, 1.00000, 0.63093);
\coordinate (F) at (0.50000, 0.63093, 1.00000);
\coordinate (X) at (0.71031, 0.71031, 0.71031);
\coordinate (P) at (0.50000, 0.76794, 0.86299);
\coordinate (Pi) at (0.39484504, 0.79675283, 0.93933188);
\coordinate (Ci) at (0.5, 1.11072107, 0.52020868);
\coordinate (V1) at (0.65111742, 0.71030992, 0.76950241);
\coordinate (V2) at (0.5       , 0.71030992, 0.92061984);
\coordinate (V3) at (0.5       , 0.63092975, 1.        );
\coordinate (V3-1rho) at (0.55919249, 0.63092975, 0.94080751);
\coordinate (V3-2rho) at (0.61838499, 0.63092975, 0.88161501);
\coordinate (V3-3rho) at (0.67757748, 0.63092975, 0.82242252);
\coordinate (V3-4rho) at (0.73676997, 0.63092975, 0.76323003);
\coordinate (V3-xrho) at (0.53906705, 0.63092975, 0.96093295);

\coordinate (rho) at (0.65111742, 0.71030992, 0.76950241);
\coordinate (rholong) at (0.41434745, 0.71030992, 1.00627238);
\coordinate (rhoarrow) at (0.44080751, 0.63092975, 1.05919249);

% local basis
\coordinate (X1) at (1,-1,0);
\coordinate (Y1) at (-1/1.732,-1/1.732,2/1.732);

% Zone subdivision
\draw [dotted] ($(A)!(O)!(B)$) -- (O);
\draw [dotted] ($(B)!(O)!(D)$) -- (O);
\draw [dotted] ($(D)!(O)!(F)$) -- (O);
\draw [dotted] ($(F)!(O)!(E)$) -- (O);
\draw [dotted] ($(E)!(O)!(C)$) -- (O);
\draw [dotted] ($(C)!(O)!(A)$) -- (O);

%%
%%
%% Drawing the interior
%%
\fill[facet] (F) -- (D) -- (B) -- (A) -- (C) -- (E) -- cycle {};
%%
%%
%% Drawing edges
%%
\draw[edge] (A) -- (B);
\draw[edge] (A) -- (C);
\draw[edge] (B) -- (D);
\draw[edge] (C) -- (E);
\draw[edge] (D) -- (F);
\draw[edge] (E) -- (F);

% objectives
\draw[->,blue] (V3) -- (rhoarrow);
% \draw[->,red] ($O+3*rho$) -- ($3*rho$);
% \draw (O) circle [x radius=0.1, y radius=0.1];
\draw[thick, red] (0.65111742, 0.71030992, 0.76950241) [x={(X1)},y={(Y1)}] circle (0.073);
\draw[thick, red] (0.65111742, 0.71030992, 0.76950241) [x={(X1)},y={(Y1)}] circle (0.15);
\draw[thick, red] (0.65111742, 0.71030992, 0.76950241) [x={(X1)},y={(Y1)}] circle (0.20);

\draw [extended line,blue] ($(O)!(V3)!(rholong)$) -- (V3);
% \draw [extended line,blue] ($(O)!(V3-1rho)!(rholong)$) -- (V3-1rho);
\draw [extended line,blue] ($(O)!(V3-2rho)!(rholong)$) -- (V3-2rho);
% \draw [extended line,blue] ($(O)!(V3-3rho)!(rholong)$) -- (V3-3rho);
\draw [extended line,blue] ($(O)!(V3-4rho)!(rholong)$) -- (V3-4rho);
\draw [extended line,blue] ($(O)!(V3-xrho)!(rholong)$) -- (V3-xrho);

% Pareto-front
\draw[-,green,line width=0.5mm] (V1) --  (V2) node[right,black] {$\vv_1$} -- (V3) node[right, black] {$\vv_2$};
\filldraw[black] (V1) circle (0.3pt) node[left, black] {$\vv_0$};
\filldraw[black] (V2) circle (0.3pt);
% \filldraw[black] (rholong) circle (0.3pt) node[left, black] {$\rho_long$};

% \draw[->,red, thick] (A) -- (Pi);
% \draw[->,red, thick] (F) -- (Ci);
% \filldraw [black] (X) circle (0.3pt) node[right, black] {$x$};
% \filldraw [black] (P) circle (0.3pt) node[right, black] {$p_1$};
%%
%%
%% Drawing the vertices
%%
\node[vertex] at (A) {};
\node[vertex] at (B)     {};
\node[vertex] at (C)     {};
\node[vertex] at (D)     {};
\node[vertex] at (E) {};
\node[vertex] at (F) {};

\end{tikzpicture}
    \caption{The Pareto curve in a \acrshort{dcg} expohedron for $\vrho=(0.55, 0.6, 0.65)$ represented as a blue arrow.
    The point $\vv_0$ is the meritocratic target exposure, \acrshort{ie} $\frac{\norm{\vgamma}_1}{\norm{\vrho}_1}\vrho$.
    The red circles are equi-unfairness curves and the blue lines are equi-utility lines.
    The green line segments form the Pareto-curve in the expohedron.
    The dotted grey lines materialize the subdivision of the expohedron into order-preserving zones.}
    % The Pareto curve always lies within an order-preserving zone.}
    \label{fig:pareto_expo}
\end{figure}
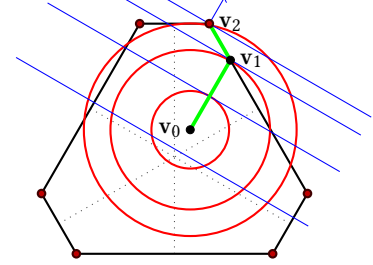

It is now easy to see that amongst all points on an equi-utility hyperplane, the point that minimizes unfairness is the one at which an equi-unfairness hypersphere is tangent, see Figure \ref{fig:pareto_expo}.
So to find the Pareto curve, we just need to start from $\vv_0=\vcEt$ and go in the direction of $\vrho$ until we intersect the border of the expohedron.
This intersection can be computed analytically using \eqref{eq:intersection}, because we stay in the same zone by moving in the direction of $\vrho$.
% For example we stay in an order preserving zone when we start from $\vv$ and go into a direction that has the same ordering as $\vv$.
% Indeed let $\vv$ and $\vx$ be two points in the expohedron.
% The intersection $\vv + \lambda(\vx-\vv)$ of the half-line $[\vv\vx)$ with the expohedron must satisfy
Indeed the intersection of a half-line $\vv+\lambda\vrho$ with the expohedron must satisfy
\begin{equation}\label{eq:intersection}
    \lambda=\arg\max_{\mu\geq0}\vv + \mu\vrho \quad\text{s.t.}\quad \vv + \mu\vrho\prec \vgamma.
\end{equation}
% The internal order of $\vv + \lambda\vrho$ does not change for $\lambda\geq0$, because $\vv$ and $\vrho$ have the same internal order.
The elements of $\vv + \lambda\vrho$ stay in the same order for $\lambda\geq0$, because $\vv$ and $\vrho$ are in the same zone.
% , it follows after some algebraic manipulations that
It follows after some algebraic manipulations that
\begin{equation}\label{eq:intersection}
    \lambda = \min_k \left\{\frac{G_k - V_k}{D_k} | D_k < 0\right\},
\end{equation}
where $G_k = \sum_{i=1}^k \vgamma^\uparrow_{i}$, $V_k = \sum_{i=1}^k \vv^\uparrow_{i}$, $D_k = \sum_{i=1}^k (\vx-\vv)^\uparrow_{i}$, and the arrow $\uparrow$ indicates a vector sorted from smallest to greatest element.

Once we are at the intersection, we can apply a similar reasoning to get the next segment of the Pareto curve.
The intersection of a hypersphere with the affine subspace containing the current face is another hypersphere of lower dimension.
Similarly the intersection of a utility hyperplane with the affine subspace containing the current face is itself an affine subspace of lower dimension.
Thus we need to go in the direction of the projection of $\vrho$ on the affine subspace containing the current face, until we meet the border of the expohedron again in a face of lower dimension.
% In order to project $\vrho$ on a face of lower dimension, we just need to find the additional normal vector.
In the end we are guaranteed to end up at a point maximizing utility, which constitutes the endpoint of the Pareto curve.

\begin{algorithm}
	\caption{Pareto set identification. We assume \acrshort{wlog} that the zone $Z(\pi)$ of $\vrho,\vv^{(0)}$ is the one where the values are ordered from smallest to greatest. This corresponds to making a change of basis such that $\pi$ becomes the identity matrix. We denote $\vgamma^\uparrow$ the vector $\vgamma$ ordered from smallest to greatest element.}\label{alg:pareto}
	\begin{algorithmic}[1]
		\Procedure{\texttt{Pareto}}{\textsc{Input}: $\vgamma$, target exposure $\vv^{(0)}$, $\vrho$}
			\State $G_k \gets \sum_{i=1}^k\vgamma^\uparrow_{i}$
			\State $V_k \gets \sum_{i=1}^k \vv^{(0)}_{i}$
			\State $\cI_0 \gets \{\ndoc\}$ \Comment{Initilize the set of splits}
			\State $\vrho^{(0)} \gets \vrho - (\vrho^\top\vone)\vone/\ndoc$ \Comment{Project $\vrho$ on the hyperplane with normal vector $\vone$}
            \State $l \gets 0$
			\While{$\vrho^\top\vv^{(l)} < \vrho^\top\vgamma^\uparrow$} \Comment{While utility is not maximal}\label{l:convergence}
    			\State $D_k \gets \sum_{i=1}^k\vrho^{(l)}_{i}$
    			\State $\lambda_l \gets \min_k \left\{\frac{G_k - V_k}{D_k} | D_k < 0\right\}$
    			\State $\vv^{(l+1)} \gets \vv^{(l)} + \lambda_l \vrho^{(l)}$ \Comment{Compute the intersection}
    % 			\State $\cI_l=\{\ndoc\}$  \Comment{$\cI_l$ is the set of splits identifying the current face.}
				\State $V_k \gets \sum_{i=1}^k\vv^{(l+1)}_{i}$
				\State $\cI_{l+1}=\{i_1,\hdots,i_{l+1}\} \gets \texttt{which}(V_k == G_k)$ \Comment{$\cI_{l+1}$ is the set of splits identifying the current face.}
				\State $\{i_j\} \gets \cI_{l+1}\setminus\cI_{l}$ \Comment{Identify the new split}
				% \State $S_1 \gets \sum_{m=i_{j-1}}^{i_j} \vgamma_m$
				% \State $S_2 \gets \sum_{m=i_{j}+1}^{i_{j+1}} \vgamma_m$
				\State $\psi \gets \sum_{m=i_{j-1}}^{i_j} \vgamma^\uparrow_m/\sum_{m=i_{j}+1}^{i_{j+1}} \vgamma^\uparrow_m$
				\State $\vnu \gets (0,\hdots,0,\underbrace{1,\hdots,1}_{i_{j-1},\hdots,i_j},\underbrace{-\psi,\hdots,-\psi}_{i_j+1,\hdots,i_{j+1}},0,\hdots,0)^\top$ \Comment The new normal vector to the face that was just intersected.
				\State $\vrho^{(l+1)} \gets \vrho^{(l)} - [(\vrho^{(l)})^\top\vnu]\vnu/\norm{\vnu}_2^2$  \Comment{Project $\vrho$ on the new face}
				\State $l\gets l + 1$
			\EndWhile
		\EndProcedure,~\textsc{Output}: a sequence of at most $\ndoc$ points $(\vv^{(l)})_{l\in\seq{0}{\ndoc-1}}$ that defines the Pareto curve as the union of the line segments connecting these points.
	\end{algorithmic}
\end{algorithm}

\begin{theo}
	Algorithm \ref{alg:pareto} returns the Pareto front for the multi-objective optimization problem \eqref{eq:pareto_moo} and has complexity $O(\ndoc^2\log(\ndoc))$.
\end{theo}

Indeed the while loop is done at most $\ndoc-1$ times and in each loop the most complex operation is the sorting operation with complexity $O(\ndoc\log(\ndoc))$.

\section{Balanced words}\label{sec:billiard}
    % Now that we have described how to obtain a Pareto-optimal solution to our MOO problem in the form of a distribution over rankings, we focus on the problem of how to deliver them as a finite sequence.
A standard approach for delivering a sequence of rankings while respecting certain desired proportions as much as possible, is to sample from a categorical distribution. %(\acrshort{aka} multinoulli) distribution.
For instance Singh and Joachims \cite{singh_fairness_2018} do a \acrshort{bvn} decomposition, then randomly sample from the obtained distribution.
% In the same vein, but with a ranking distribution expressed in the form of a \acrfull{pl} distribution, Oosterhuis \cite{oosterhuis_computationally_2021} optimizes the parameters of this \acrshort{pl} distribution \cite{plackett_analysis_1975} and directly samples the sequence of rankings from it.
In the same vein, Oosterhuis \cite{oosterhuis_computationally_2021} optimizes the parameters of a \acrfull{pl} distribution \cite{plackett_analysis_1975} and directly samples the sequence of rankings from it.

In this section we argue that there is a better way of delivering rankings from a distribution that does not involve stochastic sampling.
For example suppose we do coin flips and we have determined that heads and tails should be delivered each 50\% of the time.
If we do ten coin flips (\acrshort{ie} stochastic sampling), the probability that we get 5 heads and 5 tails (\acrshort{ie} that we are fair) is $252/1024\approx 1/4$.
Therefore a better policy is to manually take the coin and lay it alternatively once on the head once on the tail, and so on.

This intuition can be generalized to non-uniform distributions over more than two possible values by using $m$-\emph{balanced words} \cite{vuillon_balanced_2003}.
When expressed in the terms of our problem, a generator of $m$-\emph{balanced words} produces a sequence of rankings such that, in any pair of sub-strings with identical length, the frequency of any ranking differs at most by $m$.
Such a sequence is called a word and a ranking corresponds to a letter in this word.
In other words, this generator guarantees that the generated sequence delivers the rankings with proportions as close as possible to the target ones.
An important theoretical fact is that the best achievable $m$ is at most the number of unique letters (\acrshort{ie} rankings) minus 1 \cite{sano_m-balanced_2004}.
% It has been shown \cite{sano_m-balanced_2004} that when the number of letters is greater than $2$, a balanced word does not exist for every density of letters.
% However there always exists an $m$-balanced word with $m$ equal to the number of letters minus 1 \cite{sano_m-balanced_2004}.
Another advantage of this approach with respect to a \acrshort{bvn} decomposition then becomes apparent: The fact that our distribution is over at most $\ndoc$ distinct rankings instead of $(\ndoc-1)^2 + 1$ means that we are able to deliver our distribution in a much more balanced way than with a \acrshort{bvn} decomposition, \acrshort{ie} with $m=\ndoc-1$ instead of $m=(\ndoc-1)^2$.
An algorithm capable of efficiently generating $m$-balanced words, given a certain distribution of letters is given in Algorithm 1 of \cite{sano_m-balanced_2004} and in Appendix \ref{app:bw}.
This generator is equivalent to the well-known
\emph{Stride Scheduling} algorithm, used to generate fair sequences in resource (\acrshort{cpu}) management for concurrent processes \cite{waldspurger_stride_1995}.

% As an alternative to random sampling, we propose to use such an $m$-balanced word generator to deliver the ranking sequences corresponding to the Pareto-optimal solution and we empirically demonstrate its advantage \acrlong{wrt} sampling in section \ref{sec:experiments}.

\section{Experiments}\label{sec:experiments}
    In order to empirically verify our claims, we perform experiments on both synthetic data and on publicly available datasets.
Our source code is available on github\footnote{\url{https://github.com/tillkletti/expohedron}}.
We executed the computations on a laptop with an Intel\textregistered Core\texttrademark i7-8650U CPU @ 1.90GHz processor.
% We implemented our algorithms in python, the source code with a tutorial will be made available at publication time.

% However we noticed that on very large queries (typically more than 100 items) our implementation becomes sensitive to floating point precision errors, when using the bisection method. In case of such an error the implementation may be unable to

% \paragraph{\acrshort{trec}}
We use the \acrshort{trec}2020 Fairness Track evaluation queries and items \cite{trec-fair-ranking-2019} for which we computed relevance probabilities ourselves.
The computed values as well as details about the method we used to compute them will be made available in our github repository as they are not of primary importance for this paper.
% The computed relevances and the estimation method are available in the github repository, as they are not of great importance to the paper.
% \paragraph{\acrshort{mslr}}
We also use the \acrshort{mslr} dataset \cite{DBLP:journals/corr/QinL13} for which we used the ground truth relevances graded from $0$ (worst grade) to $4$ (best grade), in order to check the influence of having discrete-valued relevance values.
We normalize them to the range $[0,1]$ by dividing them by $4$.
% but of course it would be possible to compute the merit in some other way.
% The relevance vector is denoted $\vrho$.

% For both datasets, we consider the relevances to be the merit of the items.
For both datasets we restrain ourselves to queries having fewer than $100$ items, because our \acrshort{lp} baseline would take too much time on these queries and because of some numerical issues in our implementation, discussed in more detail in Appendix \ref{app:implementation}.
Furthermore we eliminate uninteresting queries having only one document and queries for which all relevance values are equal.
% Indeed for very large queries, our implementation at its current state is prone to floating point precision issues.
% Those are implementation issues that can be corrected and do not compromise the proof of concept.
This leaves 795 amongst 867 queries for \acrshort{mslr} and 198 amongst 200 queries for \acrshort{trec}.

\subsection{Metrics and baselines}
\label{ssec:Metrics}
For the exposure model we use the exposure vector $\vgamma$ of \acrshort{dcg} whose $\ixrank^\text{th}$ element is $1/\logtwo(\ixrank+2)$.
The metric measuring the utility is \acrshort{ndcg}, which divides the \acrshort{dcg} by the "ideal" \acrshort{dcg} obtained with \acrshort{prp} rankings only.
To aggregate results over several queries, we compute the arithmetic mean of the \acrshort{ndcg}s.
This indicates for a given ranking policy, what percentage of the maximal utility we achieve on average.

We consider the relevance values to be the merit of the items.
By default, we define the target exposure as $\frac{\norm{\vgamma}_1}{\norm{\vrho}_1}\vrho$, so that the sum of exposures has the right scale.
When the target exposure happens not to be inside the expohedron, this definition leads to infeasible target exposures.
% In this case there are two possibilities: We can project the target exposure onto the expohedron with the $\ell^2$ norm, a possibility we plan to develop in future work.
% In that case, we make the target feasible by choosing the smallest value $b\in\R_+$ such that $\vcEt \defeq (1-b)\frac{\norm{\vgamma}_1}{\norm{\vrho}_1}\vrho + b\frac{\norm{\vgamma}_1}{\ndoc}\vone$ becomes feasible.
In that case, we add a constant to the merits of all documents until the target becomes feasible, by choosing the smallest value $b\in\R_+$ such that $\vcEt \defeq (1-b)\frac{\norm{\vgamma}_1}{\norm{\vrho}_1}\vrho + b\frac{\norm{\vgamma}_1}{\ndoc}\vone \prec\vgamma$.
This can be efficiently computed using \eqref{eq:intersection}, since it corresponds to finding the intersection of the line segment $\left[\frac{\norm{\vgamma}_1}{\norm{\vrho}_1}\vrho,\frac{\norm{\vgamma}_1}{\ndoc}\vone\right]$ with the border of the expohedron.
% In other words we add a constant to the merits of all documents until the target becomes feasible.
% This is the choice we made in the experiments laid out here.
% From here one we suppose we have a feasible target.
% The metric measuring unfairness is $\norm{\vcE-\vcEt}_2$.
% It is a distance to the target exposure.
To aggregate results over several queries, we average a \emph{normalized unfairness}: $\frac{\norm{\vcE-\vcEt}_2}{\norm{\vgamma}_1}$ to bring the exposures to the same scale.
% Another possibility would have been to compute the mean ponderated by the number of documents affected by the unfairness. But we do not consider this other possibility.
We evaluate our own method and 4 baselines, in their ability to solve our \acrshort{moo} problem, both in terms of effectiveness and efficiency:
\begin{enumerate}
    \item Our own \acrfull{expo} method with our own Carathéodory decomposition (\acrshort{gls}) delivered using an $m$-balanced word (\acrshort{bw}) generator as described in \cite{sano_m-balanced_2004}.
    In a variant (expo end), we compute only the fairness endpoint of the Pareto-front.
    \item A \acrfull{pl} distribution \cite{plackett_analysis_1975} with parameters $\vrho$ and with a varying temperature parameter $\temperature$.
    The temperature parameter controls the interpolation between \acrshort{prp} rankings and a uniform distribution over all rankings.
    \item A \acrfull{lp} approach as proposed by \cite{singh_fairness_2018}.
    There is no trade-off between relevance and fairness with this approach; only an endpoint of the Pareto front is found, corresponding to zero unfairness, called \emph{fairness endpoint}.
    We use the \texttt{cvxopt} solver \cite{noauthor_cvxoptcvxopt_2021} to solve the \acrshort{lp} and we use a \acrshort{bvn} decomposition implemented in \cite{trabucco_brandontrabuccobvn_2021}.
    \item A \acrfull{qp} that finds bistochastic matrices maximizing, for varying $\alpha$, the scalarized \acrshort{moo}
    \begin{equation}\label{eq:scalarization}
        \max_{\pi\in\Conv(\cS_\ndoc)}\alpha\vrho^\top\pi\vgamma - (1-\alpha)\norm{\pi\vgamma-\vcEt}_2^2.
    \end{equation}
    \item A \acrfull{ctrl} as proposed by \cite{thonet_multi-grouping_2020} with parameter $K=1$ and for varying gain, hoping to find approximately Pareto-optimal solutions.
    This controller computes at each time-step whether an item is disadvantaged and tries to compensate this in future rankings by increasing its relevance with a small bonus.
    Then a \acrshort{prp} ranking is delivered using the modified relevances.
\end{enumerate}

\subsection{Results}

\paragraph{How does the runtime of our method and of the baselines vary when the number of items increases ?}
In order to fully control the number of items, we use synthetic data for this research question.
For each $\ndoc\in\seq{2}{100}$, we sample $100$ random relevance vectors whose elements are uniformly independently distributed in $[0,1]$.
For each of those $\ndoc$, we compute the full Pareto front (expo) in the expohedron.
We compute the bistochastic matrices corresponding to the fairness endpoint of the point using \acrshort{lp} and compute its \acrshort{bvn} decomposition.
For a fair comparison with \acrshort{lp}, we derive the fairness endpoint for each $\ndoc$ using our \acrshort{expo} method (expo end), and its Carathéodory decomposition with Algorithm \ref{alg:gls} (\acrshort{gls}).

\begin{figure}
    \centering
    \input{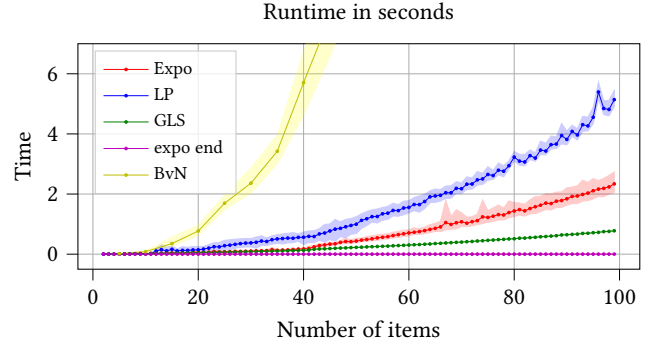}
    \vspace{-\baselineskip}
    \caption{Runtime of different algorithmic modules as a function of the number of items $\ndoc$. For each evaluated $\ndoc$, 100 random relevance vectors are sampled uniformly and independently.
    The shaded areas correspond to $95\%$ quantiles.
    The yellow curve corresponding to the \acrshort{bvn} leaves the frame, reaching 70 seconds by $\ndoc=100$. For a global comparison, the sum of the expo end and \acrshort{gls} curves should be compared with the sum of the \acrshort{lp} and \acrshort{bvn} curves.}
    \vspace{-\baselineskip}
    \label{fig:runtime}
\end{figure}

The average times the different algorithmic components take are reported in Figure \ref{fig:runtime}.
It appears that the \acrshort{bvn} decomposition is particularly slow, while the computation of the fairness endpoint using our \acrshort{expo} method is almost instantaneous.
In particular it appears that the advantage of our method (fairness endpoint computation + \acrshort{gls} decomposition) \acrshort{wrt} an \acrshort{lp} approach (including the \acrshort{bvn} decomposition) increases with increasing number of items.

% Figure \ref{fig:runtime} illustrates that the theoretical complexity does indeed correspond to the actual runtime of our method.
% Furthermore it shows how much more efficient our method is than a linear programming approach.
% Indeed using our method, one can compute the whole Pareto-frontier in less time than it takes a linear program to compute merely one endpoint of the frontier.

\paragraph{How far are the baseline methods from complete Pareto optimality?}
Among our baselines, \acrshort{qp} is guaranteed to produce Pareto-optimal points, provided that the associated \acrshort{qp} solver has sufficient precision.
For both the \acrshort{trec} and the \acrshort{mslr} datasets, we generate points of all Pareto-fronts (one front per query) by varying the trade-off parameter $\alpha$ in \eqref{eq:scalarization} between $0$ and $1$.
% For both the \acrshort{trec} and the \acrshort{mslr} dataset we compute all Pareto-fronts using our \acrshort{expo} method.
% For both the \acrshort{trec} and the \acrshort{mslr} dataset we compute all Pareto-fronts using the \acrshort{qp} method by varying the trade-off parameter $\alpha$ between $0$ and $1$.
% $\{0, 0.5, 0.15, 0.1, 0.2, 0.3, 0.4, 0.5, 0.6, 0.7, 0.8, 0.9, 0.95, 0.98, 0.99, 0.999\}$ for \acrshort{mslr} and within $\{0, 0.15, 0.2, 0.25, 0.3, 0.35, 0.4, 0.45, 0.5, 0.55, 0.6, 0.65, 0.7, 0.75, 0.8, 0.85, 0.9, 0.95, 1\}$ for \acrshort{trec}.
% Since we cannot directly aggregate Pareto fronts across queries, we need a way to parametrize the Pareto fronts.
% For each of these trade-off parameters, we compute the corresponding point on the Pareto-front found by the \acrshort{expo} method.
% By doing so we are guaranteed to find the whole Pareto front, because the Pareto front is always convex.
To plot utility-unfairness points corresponding to the \acrshort{pl} approach, we  compute for each query the performance of \acrshort{pl} after $T=1\,000$ rankings, with the temperature parameter $\temperature$ varying in $[1\e{-3}, 1]$ for \acrshort{trec} and in $[1\e{-5}, 10]$ for \acrshort{mslr}. The performances corresponding to the same temperature are aggregated over the different queries with the aggregation method described in \ref{ssec:Metrics}.

%in $\{1E-05, 0.001, 0.01, 0.015, 0.02, 0.03, 0.04, 0.05, 0.06, 0.07, 0.08, 0.09, 0.1, 0.15, 0.2, 0.3, 0.5, 1, 10\}$.
For \acrshort{ctrl}, we generate points in the same utility-unfairness space by varying the gain within $[0,1\,000]$ for \acrshort{trec} and within $[0,10\,000]$ for \acrshort{mslr}. The performances corresponding to the same gain are aggregated over the different queries to produce the plotted points.

Our  \acrshort{expo} method does not produce a parametrized set of utility-unfairness solutions.
Knowing that our \acrshort{expo} method should theoretically produce the same solutions as \acrshort{qp} for some value of $\alpha$, we check that every solution of the \acrshort{qp}, expressed as an exposure vector after multiplying the optimal bistochastic matrix by the $\vgamma$ vector, corresponds to a point of the \acrshort{expo}-derived Pareto-front and associate the corresponding $\alpha$ to the latter. % too long sentence
We are then able to aggregate the performances over the different queries, by considering the solutions associated to the same $\alpha$.
%So for each method (\acrshort{expo},\acrshort{qp}, \acrshort{pl}, \acrshort{ctrl}) we have a set of ranking policies parametrized by either $\alpha$ for \acrshort{qp} and \acrshort{expo}, by $\temperature$ for \acrshort{pl} and by the gain for \acrshort{ctrl}.
%Each (method, parameter) pair defines a ranking policy.
%For each ranking policy, we compute the average \acrshort{ndcg} and the average normalized unfairness.

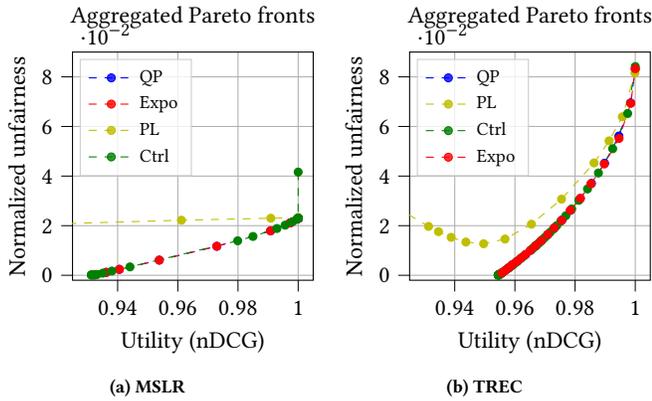
\begin{figure}
    \centering
    \begin{subfigure}[b]{0.22\textwidth}
        \centering
        % This file was created by tikzplotlib v0.9.9.
\begin{tikzpicture}

\definecolor{color0}{rgb}{0.75,0.75,0}

\begin{axis}[
legend cell align={left},
legend style={
  fill opacity=0.8,
  draw opacity=1,
  text opacity=1,
  at={(0.03, 0.97)},
  anchor=north west,
  draw=white!80!black,
  font=\footnotesize
},
tick align=outside,
tick pos=left,
title={Aggregated Pareto fronts},
x grid style={white!69.0196078431373!black},
xlabel={Utility (\acrshort{ndcg})},
xmajorgrids,
xmin=0.925, xmax=1.005,
xtick style={color=black},
y grid style={white!69.0196078431373!black},
ylabel={Normalized unfairness},
ymajorgrids,
ymin=-0.002, ymax=0.09,
ytick style={color=black},
y=33cm,
x=40cm
]
\addplot [blue, dashed, mark=*, mark size=1.5, mark options={solid}]
table {%
0.93230413312259 0.000124311093099615
0.936242050455952 0.00118531810157654
0.940560129048254 0.00238840796509496
0.95382432875407 0.00610375629549414
0.972963890194064 0.0116917693506719
0.990837151502489 0.0179285111706243
0.997303853027348 0.0210946505732291
0.999584640795482 0.0226207337810346
0.999999846981294 0.0229960237394891
0.999999877713033 0.0229960511274046
0.999999881330125 0.0229960518147879
0.999999886606983 0.0229960519756977
0.999999889808484 0.0229960515779804
0.999999892493329 0.0229960521354714
0.999999881094695 0.022996051340407
0.999999891177292 0.0229960514698894
};
\addlegendentry{QP}
\addplot [red, dashed, mark=*, mark size=1.5, mark options={solid}]
table {%
0.93210141006647 0.000123668805868048
0.936245523587871 0.00118533647068309
0.940561491212506 0.00238840693943098
0.953825051247946 0.00610381833283593
0.972964762852696 0.0116919655869483
0.990837366799767 0.0179285344114577
0.997303992296376 0.0210946493918019
0.99958475302318 0.0226207259149451
0.999999935629752 0.022996013913062
0.999999960398164 0.0229960404099735
0.999999962320494 0.0229960407001822
0.999999965380584 0.0229960404857053
0.999999966899121 0.0229960402963423
0.999999967108772 0.022996041313904
0.999999965178586 0.0229960389689969
0.999999967270351 0.0229960397789924
};
\addlegendentry{Expo}
\addplot [color0, dashed, mark=*, mark size=1.5, mark options={solid}]
table {%
1 0.0232
0.9909 0.023
0.9612 0.0222
0.9245 0.0209
0.8884 0.0199
0.8558 0.0196
0.8272 0.02
0.803 0.0207
0.7822 0.0218
0.7651 0.023
0.7092 0.029
0.6809 0.0329
0.6319 0.0399
0.6095 0.0431
};
\addlegendentry{PL}
\addplot [green!50!black, dashed, mark=*, mark size=1.5, mark options={solid}]
table {%
1 0.041609047186644
1 0.023001124946986
1 0.023001124946986
1 0.023001124946986
1 0.023001124946986
0.99976643922626 0.022781472863945
0.99813702583695 0.021599788886346
0.995662635532441 0.020193425945454
0.99293259193732 0.018860862835429
0.984982435140098 0.015655325843637
0.979903465735621 0.013903441594312
0.944101526558415 0.003381806922641
0.938082418719864 0.001701618493452
0.93505181918935 0.000857751343663
0.933227281791364 0.000351437531482
0.932616419541 0.000183564620248
0.931549484400411 9.76e-05
0.931297346759851 0.000121692052253
0.931272752911352 0.000124341044467
0.931268613126361 0.000124787227681
};
\addlegendentry{Ctrl}
\end{axis}

\end{tikzpicture}
        \vspace{-\baselineskip}
        \caption{\acrshort{mslr}}
    \label{fig:pareto_mslr}
    \end{subfigure}
    \hfill
    \begin{subfigure}[b]{0.22\textwidth}
        \centering
        % This file was created by tikzplotlib v0.9.9.
\begin{tikzpicture}

\definecolor{color0}{rgb}{0.75,0.75,0}

\begin{axis}[
legend cell align={left},
legend style={
  fill opacity=0.8,
  draw opacity=1,
  text opacity=1,
  at={(0.03, 0.97)},
  anchor=north west,
  draw=white!80!black,
  font=\footnotesize
},
tick align=outside,
tick pos=left,
title={Aggregated Pareto fronts},
unbounded coords=jump,
x grid style={white!69.0196078431373!black},
xlabel={Utility (\acrshort{ndcg})},
xmajorgrids,
xmin=0.925, xmax=1.005,
xtick style={color=black},
y grid style={white!69.0196078431373!black},
ylabel={Normalized unfairness},
ymajorgrids,
ymin=-0.002, ymax=0.09,
ytick style={color=black},
y=33cm,
x=40cm
]
\addplot [blue, dashed, mark=*, mark size=1.5, mark options={solid}]
table {%
0.954464121154926 8.41573922394198e-05
0.955629034066181 0.000948635869849571
0.956786197944614 0.00195938228833657
0.958002952727278 0.00304877068584849
0.959293071736063 0.00423358146896119
0.960643968129925 0.0055072276680924
0.962066428897372 0.00688453174686292
0.963588133891799 0.00839273012592041
0.965205444984283 0.0100486085003692
0.966943823654871 0.0118947650007588
0.968840816078799 0.0139889877698983
0.970897820458732 0.0163646256247849
0.973156590717251 0.0191045295971317
0.975626155182032 0.0222779608198165
0.978473377347513 0.0261693376257005
0.981716635364224 0.0309644319153654
0.9854422622001 0.0370496032884195
0.98978573723039 0.0451923182085734
0.994557925715105 0.0561327650454467
0.998470897164858 0.0694859058137126
0.999999805709871 0.0831273644899364
};
\addlegendentry{QP}
\addplot [color0, dashed, mark=*, mark size=1.5, mark options={solid}]
table {%
1 0.0831
0.9999 0.0816
0.9957 0.0639
0.9913 0.0542
0.9863 0.0453
0.9755 0.0307
0.9655 0.0206
0.9567 0.0146
0.9496 0.0127
0.9437 0.0134
0.9388 0.0153
0.9346 0.0176
0.9313 0.0197
0.9202 0.0277
0.914 0.0325
0.908 0.0372
0.9026 0.0414
0.8988 0.0444
0.8953 0.0471
};
\addlegendentry{PL}
\addplot [green!50!black, dashed, mark=*, mark size=1.5, mark options={solid}]
table {%
1 0.084156974521897
0.997475196140859 0.065231239767275
0.992489865613085 0.051025020126498
0.987758042002866 0.041249963113516
0.984126195815816 0.034812748492202
0.981225659478571 0.030208993602912
0.978875110252632 0.026737359541324
0.976906304817047 0.023995760924931
0.97529034001742 0.021834998947165
0.973914859140168 0.020059436884132
0.972735273424813 0.018579730495567
0.971675721359729 0.017291167333672
0.970738118644464 0.016174242491911
0.969904385210374 0.015201747725979
0.969154263529193 0.014342525679946
0.968473938904999 0.013576091081598
0.967854378636834 0.012888316553219
0.967291838115797 0.012271393365047
0.966772826658132 0.011708875304831
0.966301306664244 0.011202093329034
0.965865844546839 0.010740156231727
0.962816189861875 0.0076211615053
0.96107838105187 0.005922880642323
0.959947741227891 0.004845059361918
0.959145531975287 0.004094603127191
0.958548482311439 0.003543437864915
0.958083732064646 0.003120456259232
0.957719696116711 0.002791824414912
0.957422767041117 0.002525289401833
0.956033872451588 0.001298381711653
0.955544401442085 0.000874076760626
0.95529257257109 0.000659954492157
0.955139336493326 0.000530365166607
0.955036908197374 0.000444837703248
0.954961628011529 0.000381729021877
0.954906289781866 0.000336654720968
0.954863826616562 0.00030185411439
0.954828453524103 0.00027287465854
0.954668747380151 0.000153421744004
0.954601259283826 0.000122347781014
0.954565613782568 0.000113388541505
0.954544170837453 0.000110388807367
0.954529924157043 0.000109629085977
0.954518793455407 0.000109762226347
0.954512955158179 0.000109551212707
0.954505990658909 0.000109931679085
0.954481131747277 0.000114290949128
0.954463155308092 0.000118994011697
0.954459226340746 0.00011987900778
0.954459672436293 0.000119475679389
};
\addlegendentry{Ctrl}
\addplot [red, dashed, mark=*, mark size=1.5, mark options={solid}]
table {%
nan nan
0.955646629311943 0.000949151646801705
0.956814269813583 0.00196028624690582
0.957980836819808 0.00304196049631409
0.95905108123566 0.00421599303861156
0.96060949537996 0.00550114871124629
0.961966350525857 0.00687597906857873
0.963569250140721 0.0083982213633762
0.965072896933349 0.00999238248803626
0.96685792911751 0.011922362877971
0.968747300533738 0.0139502661579424
0.970893756266267 0.0164332657121356
0.973110195257391 0.0191286577036863
0.975524190124093 0.022107092227782
0.978408390985239 0.0261913948169566
0.981759160631063 0.031027565830661
0.985288979641509 0.0369036276459118
0.98971772324988 0.0448782886998869
0.994614736074353 0.0551912013009373
0.998518792983921 0.0693473770632032
0.999999984132986 0.0832591209449698
};
\addlegendentry{Expo}
\end{axis}

\end{tikzpicture}
        \vspace{-\baselineskip}
        \caption{\acrshort{trec}}
    \label{fig:pareto_trec}
    \end{subfigure}
    \vspace{-0.5\baselineskip}
    \caption{Aggregated Pareto fronts. The red and blue points are overlaid along the Pareto front and therefore hard to distinguish.}
    \vspace{-\baselineskip}
    \label{fig:pareto}
\end{figure}

% The results are reported in Figure \ref{fig:pareto_trec} for \acrshort{trec} and in Figure \ref{fig:pareto_mslr} for \acrshort{mslr}.
The results are reported in Figure \ref{fig:pareto}.
It appears without surprise that both \acrshort{qp} and \acrshort{expo} perform identically, while \acrshort{pl} is far from Pareto-optimality, except in the region where \acrshort{ndcg} reaches $1$.
More surprisingly \acrshort{ctrl} performs well for a heuristic approach, with its points being indistinguishable from the actual Pareto front, except for an outlier obtained with $\text{gain}=0$ for the \acrshort{mslr} dataset.
% This outlier can be explained by the fact that for discrete relevance values there are many ties,.
However, as it does not explicitly address the problem as a MOO problem and as it uses a parameter (the  gain) which controls very indirectly the utility-fairness trade-off, there is as of yet no guarantee that it will always result in near-optimal Pareto-fronts in all settings.

% \begin{figure}
%     \centering
%     % \includegraphics{}
%     \input{figures/pareto_fronts_MSLR}
%     \caption{Aggregated Pareto fronts for \acrshort{mslr} dataset}
%     \label{fig:pareto_mslr}
% \end{figure}

% \begin{figure}
%     \centering
%     % \includegraphics{}
%     \input{figures/pareto_fronts_TREC}
%     \caption{Aggregated Pareto fronts for \acrshort{trec} dataset}
%     \label{fig:pareto_trec}
% \end{figure}

\paragraph{How do the objectives evolve over time ? Does using balanced words provide an advantage \acrshort{wrt} sampling ?}

For each method we select the parameter that brings it closest to minimal unfairness.
For \acrshort{expo} this corresponds to simply choosing the fairness endpoint.
For \acrshort{pl} we select $\temperature=1$.
For \acrshort{ctrl} we select a gain of $0.6$ for \acrshort{trec} and a gain of $10$ for \acrshort{mslr}.
For \acrshort{lp} there is no parameter to set and we do not apply a \acrshort{qp}, since \acrshort{lp} does the same job more efficiently.
We aggregate the results over all queries by taking the average normalized unfairness at each time-step.
% The results are reported in Figure \ref{fig:time_evo_mslr} for \acrshort{mslr} and in Figure \ref{fig:time_evo_trec} for \acrshort{trec}.

\begin{figure}
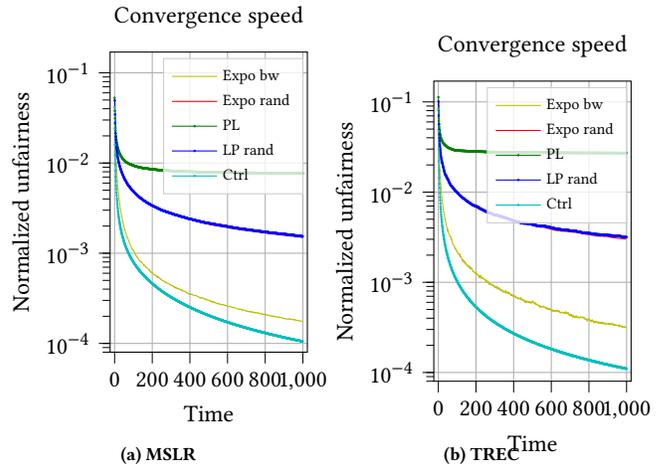

    \centering
    \begin{subfigure}[b]{0.23\textwidth}
        \centering
        \input{figures/objective_evolution_MSLR_bis}
        \vspace{-\baselineskip}
        \caption{\acrshort{mslr}}
        \label{fig:time_evo_mslr}
    \end{subfigure}
    \hfill
    \begin{subfigure}[b]{0.23\textwidth}
        \centering
        \input{figures/objective_evolution_TREC}
        \vspace{-\baselineskip}
        \caption{\acrshort{trec}}
        \label{fig:time_evo_trec}
    \end{subfigure}
    \vspace{-0.5\baselineskip}
    \caption{Average normalized unfairness as a function of the number of delivered rankings.
    The red and blue curves are almost equal and may be difficult to distinguish.}
    % \vspace{-\baselineskip}
    \label{fig:time_evo}
\end{figure}

The results are reported in Figure \ref{fig:time_evo}.
It appears that sampling from a distribution found via \acrshort{bvn} or via \acrshort{gls} has no notable effect on transient performance.
However using balanced words instead of sampling does have a beneficial effect on the speed of convergence to the target exposure.
It also appears that \acrshort{ctrl} does converge more quickly than all other methods, at least for the fairness endpoint.

% \begin{figure}
%     \centering
%     \input{figures/objective_evolution_MSLR}
%     \caption{Evolution of objective metrics with time for \acrshort{mslr}}
%     \label{fig:time_evo_mslr}
% \end{figure}

% \begin{figure}
%     \centering
%     \input{figures/objective_evolution_TREC}
%     \caption{Evolution of objective metrics with time for \acrshort{trec}}
%     \label{fig:time_evo_trec}
% \end{figure}

\paragraph{Is \acrshort{expo}~ faster than existing baselines ?}

In Table \ref{tab:runtimes} we report the time it took to compute the solutions using our \acrshort{expo} method and the baselines.
% We used the implementation available in our github repository and executed the computation on a laptop with an Intel(R) Core(TM) i7-8650U CPU @ 1.90GHz processor.
% To measure the time, we use the python function \texttt{time.time()}.
%, while performing some inexpensive tasks such as text document editing or web browsing.
It appears that \acrshort{expo} is clearly superior to the other exact methods \acrshort{lp} and \acrshort{qp} in terms of runtime.
For $T=1\,000$ rankings \acrshort{ctrl} is quicker than \acrshort{expo}.
However for $T=10\,000$ rankings, \acrshort{ctrl} takes ten times as long, whereas for the \acrshort{expo} method, the computation of the Pareto-front (or endpoint) and \acrshort{gls} need to be performed but once.
Then once a solution is found, it can be delivered very quickly using balanced words.
Thus for larger time horizons with an order of magnitude of several thousand rankings, \acrshort{expo} becomes quicker than \acrshort{ctrl} for the total runtime.

\begin{table}[h]
    % \small
    \centering
    \caption{Average runtimes in seconds for each dataset for $T=1\,000$ and $T=10\,000$ rankings.}
    % It appears that for $T=1\,000$ rankings, \acrshort{ctrl} is the fastest method.
    % In the long run however, for $T=10\,000$, \acrshort{expo} becomes the fastest method for \acrshort{trec}.
    % This happens because \acrshort{expo} has a fixed cost at the beginning, but the delivery using balanced words is very cheap, whereas \acrshort{ctrl} has a cost that increases linearly with $T$.}
    \tabcolsep=0.11cm
    % \singlespacing
    \begin{tabular}{|c|c|c|c|c|}
        \hline
        & \shortstack{\acrshort{trec}\\ $T=1\,000$} & \shortstack{\acrshort{mslr}\\ $T=1\,000$} & \shortstack{\acrshort{trec}\\ $T=10\,000$} & \shortstack{\acrshort{mslr}\\ $T=10\,000$} \\
        \hline
        \acrshort{expo} endpoint & 0.0014 & 0.0018 & 0.0014 & 0.0018\\
        \acrshort{expo} & 0.0806 & 0.0199 & 0.0806 & 0.0199\\
        \acrshort{lp} & 0.0858 & 2.1369 & 0.0858 & 2.1369\\
        \acrshort{qp} & 0.8424 & 15.2095 & 0.8424 & 15.2095\\
        \hline
        \acrshort{gls} & 0.0737 & 0.5396 & 0.0737 & 0.5396\\
        \acrshort{bvn}  & 2.2623 & 29.1706 & 2.2623 & 29.1706\\
        \hline
        \acrshort{bw} & 0.0009 & 0.0010 & 0.0094 & 0.0093\\
        rand & 0.0388 & 0.0469& 0.4060 & 0.4340\\
        \hline
        \hline
        \acrshort{expo}+\acrshort{gls}+\acrshort{bw} & 0.1552 & 0.5605 & 0.1637 & 0.5688\\
        \acrshort{lp}+\acrshort{bvn}+rand & 2.3869 & 31.3544 & 2.7541 & 31.7415\\
        \acrshort{ctrl} & 0.0245 & 0.0268 & 0.2201 & 0.2543\\
        \acrshort{pl} & 0.0145 & 0.0174 & 0.1525 & 0.1991\\
        % \acrshort{ctrl} & 0.0245 & 0.0268 & 0.2201 & 0.2543\\
        % \acrshort{pl} & 0.0145 & 0.0174 & 0.1525 & 0.1991\\
        \hline
    \end{tabular}
    % \vspace{\baselineskip}
    \label{tab:runtimes}
\end{table}

\section{Conclusion}\label{sec:conclusion}
    Our novel geometrical framework makes it possible to efficiently compute all Pareto-optimal fairness-utility amortizations for a \acrshort{pbm}.
% While a controller from \cite{thonet_multi-grouping_2020} does the job more quickly for time horizons lower than several thousands, our approach does not depend one any parameter.
Amongst the methods that are provably Pareto-optimal, our method is the overall quickest.
% The controller from \cite{thonet_multi-grouping_2020} empirically performs equally well and is quicker for time horizons lower than several thousands, but does not allow to set the desired trade-off in advance, except by an expensive grid search, because the best parameter given a desired trade-off depends on the dataset.
The controller from \cite{thonet_multi-grouping_2020} empirically performs equally well and is quicker for time horizons lower than several thousands.

In future work we plan to extend our framework to group fairness and to different exposure models that are not \acrshort{pbm}s, such as \acrlong{dbn} models \cite{chuklin_click_2015}.
% It is in principle also possible to extend our work to many different exposure models that are not \acrshort{pbm}s, but an efficient Carathéodory decomposition for those exposure models remains to be found.

% Discuss extension to other $p$-norms with $p\neq2$, in particular $p\in\{1, +\infty\}$. The case $p=+\infty$ corresponds to \emph{Rawlsian max-min fairness} (expression taken from \cite{zhu_fairness_2021}).
% Discuss possibility of better Carathéodory decompositions.
% perhaps discuss the fact that joachims have no unfairness metric

%%
%% The acknowledgments secWe give here the ption is defined using the "acks" environment
\begin{acks}
    This work has been partially supported by MIAI@Grenoble Alpes, (ANR-19-P3IA-0003).
\end{acks}

% \newpageWe give here the p
\newpage
\bibliographystyle{ACM-Reference-Format}
\balance
\bibliography{WSDM2022,biblio}
\clearpage

% \appendix

\section*{Appendices}
\renewcommand{\thesubsection}{\Alph{subsection}}

\subsection{Implementation}\label{app:implementation}
    Our algorithms \ref{alg:gls} and \ref{alg:pareto} are exact algorithms on paper, but in practice, when implemented in a machine that cannot handle real numbers, some numerical difficulties are encountered.
In this section we explain some tricks we used to tackle these difficulties and their limitations.

Consider Algorithm \ref{alg:gls}.
When doing a bisection search with a given precision on a half-line, in order to find its intersection with the border of the expohedron, the point $\vp$ that is found only approximately lies on the intersected face.
Therefore the next half-line starting at a vertex $\vv$ and passing through $\vp$ is not exactly contained in the subspace containing the face of $\vv$ and $\vp$.
The further away a point on the half-line is from $\vv$, the bigger this error will be.
In the worst case the error will be big enough so that the next intersection with the border of the expohedron is not on a face of lower dimension, thereby preventing convergence of the algorithm.
Such a problem can be avoided by projecting the approximate point $\vp$ on the hyperplane containing the face on which $\vp$ lies.
This substantially reduces the error of $\vp$ and makes it possible to avoid many convergence problems.

Our implementation still encounters convergence problems when the number of documents is high.
In that case the expohedron is a very high-dimensional polytope and some faces are smaller than the implementation's precision, which makes face-identification error-prone by the end of Algorithm \ref{alg:pareto}, when it is already close to maximum utility.
This issue can be avoided by putting a tolerance in line \ref{l:convergence} of Algorithm \ref{alg:pareto}.
Typically for our experiments, setting a tolerance of $10^{-6}$ was sufficient.

% Another difficulty which has yet to be solved arises when the number of documents is very high.
% In that case the the expohedron is a high-dimensional polytope and with some faces that are smaller than the precision with which faces are identified.

\subsection{Balanced words}\label{app:bw}
    We lay out here Algorithm \ref{alg:balanced_words} to efficiently generate $m$-balanced words \cite{sano_m-balanced_2004}.
In practice we let let the algorithm warm up for a few hundred iterations, when $\vphi$ is initialized to $\vzero$.
Otherwise the algorithm starts by delivering each ranking once, regardless of its density.
% Another more efficient possibility to avoid this, would be to initialize $\vphi_i$ to $\alpha_i$ for every $i$ in $\seq{1}{N}$.

\begin{algorithm}
	\caption{A generator of $m$-balanced words}\label{alg:balanced_words}
	\begin{algorithmic}[1]
		\Procedure{\texttt{BW}}{\textsc{Input}: a density $\alpha_1,\hdots,\alpha_N\in\R_+,~\sum_{i=1}^N\alpha_i=1$ and an alphabet $\cA$ with elements $(a_i)_{i\in\seq{1}{N}}$}
		\State $\vphi \gets \vzero\in\R^N$
		\State $t \gets 0$
		\While{\texttt{True}} \Comment{Generate the next element of the sequence}
		    \State $i^* \gets \arg\min_{i\in\seq{1}{N}}\vphi_i$
		    \State $\vx_t \gets a_{i^*}$
		    \State $\vphi_{i^*} \gets \vphi_{i^*} + \frac{1}{\alpha_{i^*}}$
		    \State $t \gets t + 1$
		\EndWhile
		\EndProcedure~\textsc{Output}: A sequence $\vx\in\cA^\N$. $\vx$ is an $N-1$-balanced word.
	\end{algorithmic}
\end{algorithm}

\end{document}